\documentstyle[prb,preprint,aps]{revtex}
\begin{document}


\title {Numerical renormalization group approach to
fluctuation exchange in the presence of electron-phonon
coupling: Pairing in the Holstein--Hubbard model}
\author{C.--H. Pao} 
\address{Department of Physics, National Chung Cheng University,\\ 
  Chia--Yi, 621 Taiwan, R.O.C.}
\author{H.--B. Sch\"uttler}
\address{Center for Simulational Physics, Department of Physics \& Astronomy,\\
 University of Georgia, Athens, GA  30602 }
\date{\today}

\maketitle

\begin{abstract}
The fluctuation exchange (FLEX) approximation is applied to
study the Holstein--Hubbard model. Due to the retarded nature
of the phonon-mediated electron-electron interaction, 
neither fast Fourier transform (FFT) nor previously developed NRG
methods for Hubbard-type purely electronic models
are applicable, while brute force
solutions are limited by the 
demands on computational time and storage which increase rapidly 
at low temperature $T$. 
Here,we describe a new numerical renormalization 
group (NRG) technique to solve the FLEX equations
efficiently. Several orders of magnitude of CPU time and storage
can be saved at low $T$ ($\sim\, 80K$). 
To test our approach, we compare our NRG results
to brute force calculations on small lattices at
elevated temperatures.
Both  $s$--wave and $d$--wave superconducting phase diagrams
are then obtained by applying the NRG approach at low $T$. 
The isotope effect for $s$--wave pairing is 
BCS--like in a realistic phonon frequency range, 
but vanishes at unphysically large phonon frequency
($\sim\,$ band width). 
For $d$--wave pairing, the isotope
exponent is negative and small compared 
to the typical observed values in 
non-optimally doped cuprates. 
\end{abstract}
%

\section{Introduction}
\label{sec:intro}

A large body of experimental evidence suggests
that several high--$T_c$ cuprate superconductors 
exhibit a pairing state of $d_{x^2-y^2}$ 
symmetry \cite{vanh95,scal95}.
In combination with the large superconducting transition
temperature of these materials,
this suggests a superconducting pairing mechanism of 
predominantly electronic origin. In particular
anti--ferromagnetic (AF) spin fluctuation (SF) exchange
has been proposed as a possible electronic candidate mechanism
\cite{bick87,bick89,bick91,yone89,moriya,pines} 
which would tend to 
give rise to $d_{x^2-y^2}$ pairing symmetry.

However, except near certain ``optimal'' doping concentrations,
many cuprates exhibit a quite noticeable
doping dependent isotope 
effect\cite{isoexp1,isoexp2,isoexp3,isoexp4}
and other pronounced, superconductivity-related 
lattice effects \cite{latteff}. 
This indicates that electron--phonon (EP) interactions 
could be important and should be included in the 
theory.\cite{schu95,pao_tbp}

In the past decade, 
conserving self-consistent field (SCF)
methods\cite{baym61} and related diagrammatic 
approaches, such as the fluctuation exchange 
(FLEX) approximation 
\cite{bick89,bick91,sere91,luo93,pao94a,pao94,lenc94,pao95}
have been used to study AF SF exchange
within the framework of microscopic correlated electron models.
Most of the FLEX-based SCF 
studies so far have been limited to Hubbard-type models with
instantaneous, local Coulomb interactions
in simple tight--binding models, and some extensions
to long-range Coulomb interactions \cite{esir97,esir98}.
For these types of model systems, both numerical renormalization 
group (NRG)\cite{pao94a} and fast Fourier transform (FFT)\cite{sere91}
methods have been developed and successfully applied to solve the FLEX 
equations efficiently on large space-time lattices in the physically 
relevant low-temperature regime. The numerical solution of the FLEX
equations is greatly simplified in instantaneous interaction models, 
due to the fact that the bare electron-electron interaction potential
is frequency independent.

When EP interactions are introduced into the model, 
already the bare electron-electron potential becomes 
explicitly frequency dependent, due the retarded nature of the
phonon-mediated interaction. In that case, the solution
of the FLEX equations requires the inversion of certain, large 
fermion frequency matrices
which, in a brute force approach, would increase
the demands on CPU time and memory consumption by several 
orders of magnitude, relative to the purely electronic models
studied so far. Also, neither FFT approaches \cite{sere91}
nor the original form of the NRG method \cite{pao94a}
can be used here to reduce the numerical effort to
manageable proportions.

Previous SCF studies of strongly correlated electron models with 
EP coupling \cite{mars90} have included renormalization of the Coulomb and phonon-mediated potential at the level of an RPA particle-hole bubble.
In this much simpler approach, one neglects the electron--electron 
exchange scattering which arises in the full FLEX approximation,
due to the Pauli exclusion principle.
This simpler RPA-based approximation \cite{mars90} then avoids the
frequency matrix inversion problem, since the latter arises only
from the exchange ladder diagrams.
This may be a good approximation when the Coulomb 
repulsion is strongly screened and the phonon frequency is small.
These are, essentially, the conditions under which Migdal's theorem is valid
\cite{migd58} and the Eliashberg theory \cite{elia60,scal69,allen_mitr} 
of phonon-mediated conventional superconductivity
is applicable. In general, these conditions may not be satisfied in 
correlated electron systems and it may be necessary to include both Coulomb 
and EP contributions to the electron exchange vertex.

In the present paper, we describe an extension of the NRG approach
which will allow us to incorporate EP interactions into a 
Hubbard-type correlated electron model 
and handle the resulting frequency matrix inversions in the FLEX equations 
efficiently. An efficient algorithm to solve this matrix inversion problem 
in the present context is also an important first step towards studying the 
next level of SCF theory, such as, for example, the parquet 
theory \cite{bick91,domi64,bick98}.
Our present treatment is limited to the case of the so--called
Holstein-Hubbard model
\cite{mars90,schu92,free95} 
where the
phonon-mediated electron-electron 
potential is without momentum dependence. 
However, when combined with recently developed real-space basis
representation approaches,\cite{esir97,esir98,hank74} 
our basic method to the frequency
matrix inversion problem will also become applicable to  
bare potentials
which are frequency {\it and} momentum dependent.

Preliminary results obtained with our
present frequency matrix NRG technique
for the $d$-wave instability of the Holstein-Hubbard
model have been reported elsewhere \cite{pao98}.
The purpose of the present paper is to give a first 
detailed account of the technique itself,
to present results for the $s$-wave instability
and for the competition between $d$- and $s$-wave pairing
in their respective parameter regimes. 

The paper is organized as follows: in Section II, 
we summarize the FLEX approximation 
for the Holstein--Hubbard model 
and define the notations used in the paper.
In Section III,
we describe our fermion frequency matrix
NRG technique in detail. 
In Section IV, we present our results
obtained with the NRG approach for the Holstein--Hubbard 
model. They include a comparison of results
for one--particle correlation functions 
obtained with the NRG and, respectively, 
by the brute force approach; 
and some applications to the 
superconducting instabilities and the isotope
effect of the Holstein--Hubbard model. 
We conclude with a brief summary in Section V.

\section{Holstein--HUbbard model in the FLEX approximation}
\label{sec:model}

We start from the simplest microscopic model, including both
an on--site Hubbard $U$ Coulomb repulsion and a local 
Holstein-type EP
coupling to an Einstein phonon branch,
The Hamiltonian of this Holstein--Hubbard model
\cite{mars90,schu92,free95}
can be written as:
\begin{eqnarray}
  H &=& -t\sum_{\langle ij\rangle\sigma}\thinspace
    \biggl [ c^{\dagger}_{i \sigma}
    c^{\vphantom{\dagger}}_{j \sigma} + HC \biggr ]\ -\
    \mu\sum_{i\sigma}\thinspace n_{i \sigma} \ +\
    U\sum_i\thinspace n_{i \uparrow} n_{i\downarrow} \nonumber\\
    & & +\ \sum_{i}\thinspace \left [{ {{ p}_{i}}^2 \over 2M}
        + {1 \over 2} K {{ u}_i}^2 \right ]\ -\
        C \sum_{i\sigma}\thinspace {u}_i
           \Big(n_{i \sigma} - {1 \over 2}\Big)\ ,
\end{eqnarray}
with a nearest neighbor hopping $t$, chemical potential $\mu$,
on--site Coulomb repulsion $U$, on--site EP coupling constant $C$,
harmonic restoring force constant $K$, and ionic oscillator mass $M$.
The $c^{\dagger}_{i \sigma}$
($c^{\vphantom{\dagger}}_{i \sigma}$) is the electron 
creation (annihilation) operator at site $i$ 
and spin $\sigma$; $n_{i \sigma}$ is the number operator;
and  ${ u}_i$ is the local ionic displacement at lattice site $i$
and $p_i=-i\hbar\partial/\partial u_i$.
The dispersionless bare phonon frequency is 
\begin{equation}
\label{eqnOmega_P}
\Omega_p\,=\, (K/M)^{1 / 2}
\end{equation}
and the phonon-mediated on--site attraction is 
\begin{equation}
\label{eqnU_P}
U_p\,=\,C^2 / K\;.
\end{equation}

The bare interaction vertices entering into the FLEX
treatment are shown in
Fig.\ref{vert}, including
the particle--hole density and magnetic vertices 
[ Figs. \ref{vert}(b) and \ref{vert}(c)]
\begin{equation}
V^D_{n_1,n_2,n_3,n_4}(i\nu_m)  = 
v^D_{n_1,n_4}(i\nu_m) 
\delta_{n_1,n_2+m} \delta_{n_3+m,n_4}\ ,
\end{equation}
where
\begin{equation}
v^D_{n_1,n_4}(i\nu_m)  = 
 2 v_p(i \nu_m)\, -\, v_p(i\omega_{n_1}-i\omega_{n_4})\, +\, U 
\end{equation}
and
\begin{equation}
V^M_{n_1,n_2,n_3,n_4}(i\nu_m)  = 
v^M_{n_1,n_4}(i\nu_m) 
\delta_{n_1,n_2+m} \delta_{n_3+m,n_4}\ ,
\end{equation}
where
\begin{equation}
v^M_{n_1,n_4}(i\nu_m)  = 
 - [ v_p(i\omega_{n_1}-i\omega_{n_4})\, +\, U] \ ,
\end{equation}
and the particle--particle singlet and triplet vertices
[ Fig. \ref{vert}(d)]
\begin{equation}
V^{S}_{n_1,n_2,n_3,n_4}(i\nu_m) = 
v^{S}_{n_1,n_4}(i\nu_m)
 \delta_{n_1,-n_2+m} \delta_{-n_3+m,n_4}\ ,
\end{equation}
where
\begin{equation}
 v^{S}_{n_1,n_4}(i\nu_m) = 
{1\over 2} [ v_p(i\omega_{n_1}-i\omega_{n_4})
\ +\ v_p(i\omega_{n_1}+i\omega_{n_4} -i\nu_m)\, +\, 2 U]
\end{equation}
and
\begin{equation}
 V^{T}_{n_1,n_2,n_3,n_4}(i\nu_m) = 
 v^{T}_{n_1,n_4}(i\nu_m)
 \delta_{n_1,-n_2+m} \delta_{-n_3+m,n_4}
\end{equation}
where
\begin{equation}
v^{T}_{n_1,n_4}(i\nu_m) = 
{1\over 2} [ v_p(i\omega_{n_1}-i\omega_{n_4})
\, -\, v_p(i\omega_{n_1}+i\omega_{n_4} -i\nu_m) ] \ .
\end{equation}
Here, the  electron-electron potential includes both the 
the Hubbard $U$ and the phonon-mediated contribution
[ Fig. \ref{vert}(a)]
\begin{equation}
v_p(i \nu_m)\ =\ -{U_p \Omega_p^2  / (\Omega_p^2 + \nu_m^2)}\ .
\end{equation}
The boson Matsubara frequencies are denoted by
$\nu_m = 2 m \pi T$ and the fermion Matsubara frequencies by
$\omega_n = (2 n +1) \pi T$ with integer $m$ and $n$.

In the FLEX approximation, the single--particle self-energy 
is then given by\cite{bick89}
\begin{eqnarray}
\label{eqnself}
\Sigma(k)
         & =&\sum_{{\bf k^\prime},i\omega_{n^\prime}}\ 
 {T\over N}  \Bigl \{ \Bigl [
   V_2(k-k^\prime; i\omega_n )\ + \
   V^{ph}(k-k^\prime; i\omega_n) \Bigr ] G(k^\prime)
\nonumber
\\
&\;&\;\;\;\;\;\;\;\;\;\;\;\;\;\ \,+
   V^{pp}(k-k^\prime; i\omega_n) G^{*}(k^\prime)
        \Bigr \} 
\end{eqnarray}
where the effective interaction potentials are given by\cite{bick89}
\begin{eqnarray}
\label{eqnv2}
V_2(q; i\omega_n ) & = &-v_p(i\nu_m)\ + \ \nonumber \\
 &&\sum_{i \omega_{n^{\prime\prime}}}
  \Bigl [ v_p(i\nu_m) + U \Bigr ] \Bigl [ 2 v_p(i\nu_m) -
   v_p(i\omega_n - i \omega_{n^{\prime\prime}} - i\nu_m) + U \Bigr ]
   \bar{\chi}_{ph} ( q; i\omega_{n^{\prime\prime}})\ , \\
V^{ph}(q; i\omega_n ) & = & \sum_{i \omega_{n^{\prime\prime}}}\ 
  \Bigl \{ {1 \over 2}
  \Bigl [ D(1+D)^{-1} - D \Bigr ]_{ n, n^{\prime\prime}}\!\! (q)\ 
     v^D_{n^{\prime\prime}, n}(i\nu_m)\ + \nonumber \\
\label{eqnvph}
  && \ \ \ \ \ \ { 3 \over 2}
  \Bigl [ M(1+M)^{-1} - M \Bigr ]_{ n, n^{\prime\prime}}\!\! (q)\ 
    v^M_{n^{\prime\prime}, n} (i\nu_m)
  \Bigr \} \ , \\
\label{eqnvpp}
V^{pp}(q; i\omega_n ) & = & -\sum_{i \omega_{n^{\prime\prime}}}
  \Bigl \{ \
  \Bigl [ S(1+S)^{-1} - S \Bigr ]_{ n, n^{\prime\prime}} \!\! (q)\ 
     v^S_{n^{\prime\prime}, n}(i\nu_m)\ + \nonumber \\
  && \ \ \ \ \ \ \ 3
  \Bigl [ T(1+T)^{-1} - T \Bigr ]_{ n, n^{\prime\prime}} \!\!(q)\ 
     v^T_{n^{\prime\prime}, n}
     (i\nu_m)
  \Bigr \}\ ,   \\
\label{eqnRmat}
 R_{n, n^{\prime\prime}} (q) & = & v^R_{n, n^{\prime\prime}}(i\nu_m)
   \ \  \Biggl \{ \begin{array}{l}
  \bar{\chi}_{ph}(q; i\omega_{n^{\prime\prime}})\  \\
  \bar{\chi}_{pp}(q; i\omega_{n^{\prime\prime}}) \end{array}
   \ \ \ \ \ \ \ {\rm for} \ \ R \ =\ \Biggl \{\begin{array}{l}
      D\ {\rm or}\ M\ ,\\
      S\ {\rm or}\ T\ .
  \end{array}
\end{eqnarray}
Here, we are using a momentum-energy vector notation
where, for fermion lines,
$k \, \equiv\, ({\bf k},i\omega_n)$ and
$k^\prime \, \equiv\, ({\bf k^\prime},i\omega_{n^\prime})$
and, for boson lines,
$q \, \equiv\, ({\bf q},i\nu_m)$.
The Green's function is
\begin{equation}
G(k)\, =\,
[ i\omega_n - \epsilon_{\bf k} - \Sigma(k) ]^{-1}
\ ,
\label{eqndyson}
\end{equation}
and the tight binding band
\begin{equation}
\epsilon_{\bf k}\, =\, -2t (\cos \, {\bf k}_x + \cos \, {\bf k}_y)
- \mu
\ .
\end{equation}
The bare particle--hole and particle--particle fluctuation functions
are then defined as:
\begin{eqnarray}
\bar{\chi}_{ph}(q; i\omega_n) & = &
   -{T \over N} \sum_{\bf k} G(k+q) G(k) \ ,
\label{eqnxph}
\\
\bar{\chi}_{pp}(q; i\omega_n) & = &\ \  
{T \over N} \sum_{\bf k} G(k+q) G(-k) 
\label{eqnxpp}\ ,
\end{eqnarray}
{ without} summation over the fermion Matsubara frequency $i\omega_n$. 

Eqs.~(\ref{eqnself}--\ref{eqnxpp}) are solved iteratively.
The iteration proceeds as follows: 
(1) choose the temperature
and either a fixed electron concentration $\langle n \rangle$ 
or a fixed chemical potential $\mu$;
(2) guess an initial 
self-energy $\Sigma(k)$ at this temperature and electron concentration,
{\it e.g.},  $\Sigma\,\equiv\,0$;
(3) calculate the Green's function $G(k)$ from Eq.~(\ref{eqndyson}); 
(4) calculate the 
bare particle--hole and particle--particle fluctuation functions
by equations (\ref{eqnxph}) and (\ref{eqnxpp}) using the 
Green's function obtained in step (3); 
(5) evaluate the matrices
$R$ in all four channels and the effective potentials $V_2$, 
$V^{ph}$, and $V^{pp}$; 
(6) update the self-energy by Eq.~(\ref{eqnself}); 
(7) using the updated self-energy $\Sigma$ from step (6) 
as  input, go back to step (3) to calculate 
an updated Green's function $G$.  
The iterative cycle, consisting
of steps (3) through (7), is then repeated until a
converged self-energy is obtained. If $\langle n\rangle$
is to be fixed to a given input value, then the chemical
potential $\mu$ must be ajdusted accordingly during the
self-consistent calculation.

Because of the retarded nature
of $v_p(i \nu_m)$ the
bare vertices in Fig.1(b-d) depend
explicitly on the internal frequency
transfer. As a consequence, a frequency matrix inversion 
is necessary to evaluate the 
fluctuation potentials, $V^{ph}$ and $V^{pp}$,
in Eqs.~(\ref{eqnvph}) and (\ref{eqnvpp}).
In a brute force approach, this matrix dimension
can become as large as $500^2$ to $1000^2$ (the size
of the entire fermion Matsubara frequency set)
near the transition temperature. Furthermore, for each iteration,
the number of such matrix inversions to be carried out
is about the same as the number of boson Matsubara frequencies 
times the size of the momentum grid.
At the space-time lattice sizes required to study
the physically interesting low-temperature regime,
it is therefore not possible to carry 
out such a brute force calculation with currently
available computing resources.
Recent FFT and numerical NRG techniques, 
developed for the pure Hubbard FLEX
equations are also not directly applicable.
In the next section,  we will describe 
a generalized ``fermion frequency matrix'' NRG method 
to handle the numerics efficiently.

After a convergent self-energy is obtained at a fixed 
temperature and electron filling, the search for
pairing instabilities requires the calculation of the maximal 
eigenvalue of the pairing kernel,
$\lambda(T)$, as a function of temperature $T$, from
\begin{equation} \label{eqnlambda}
\lambda(T) \phi(k) \ =\ 
-{T\over N} \sum_{{\bf k^\prime},i\omega_{n^\prime}} 
V_{\rm pair}
   (k,k^\prime;T) G(k^\prime) G(-k^\prime) \phi(k^\prime) \ ,
\end{equation}
where the pairing potential $ V_{\rm pair}$ is \cite{bick89}:
\begin{eqnarray}
\label{eqnvpair}
V_{\rm pair}(k,k^\prime;T)\ =\ 
   { 1 \over 2}v_{p}(i \omega_n - i\omega_{n^\prime}) 
& -& {1 \over 2} \bigl [ D (1 + D)^{-1} \bigr ]_{n,n^\prime} (k-k^\prime) 
      v^D_{n^\prime,-n} (k-k^\prime) \nonumber \\
& +& {3 \over 2} \bigl [ M (1 + M)^{-1} \bigr ]_{n,n^\prime} (k-k^\prime) 
      v^M_{n^\prime,-n} (k-k^\prime)\ .
\end{eqnarray}
The instability is reached when $\lambda(T)$ approaches unity, {\it i.e.},
\begin{equation}
\lambda(T) \ \rightarrow \ 1\ \ \ \Rightarrow\ \ \   T\ \rightarrow\ T_c\ .
\end{equation}  

\section{Numerical renormalization group 
approach for the Holstein--Hubbard model}
\label{sec:rg}

A numerical NRG method has been successfully applied to the
FLEX equations of the Hubbard 
model \cite{pao94a}, which has a frequency- and momentum-indepedent
bare interaction, the on--site Coulomb $U$.
In this case the matrix inversion
in equations (\ref{eqnvph}) and (\ref{eqnvpp}) can be carried out
analytically and the NRG operations are greatly simplified \cite{pao94a}. 
%
Due to the frequency dependence of the phonon-mediated interaction $v_p$,
we now have to construct a more general NRG operation in which the
frequency dependence of the bare interaction is taken into account.
The detailed procedure will now be described.

The NRG evaluation of the self-energy follows closely
the original NRG approach described in Ref.~\onlinecite{pao94a}.
We will largely adopt the notation and terminology introduced therein.
We are implementing a pure ``frequency NRG''
(in the sense of Ref.~\onlinecite{pao94a}). That is,
the grid of momentum points ${\bf k}$ is chosen from the outset
to be dense enough for the lowest temperatures to be reached
and remains constant throughout all NRG steps;
only the Matsubara frequency grids ($i\omega_n$, $i\nu_m$) 
change from one NRG step to the next.

The basic assumption underlying the NRG approach is that

(1) quantities such as the $\Sigma(k)$ and $G(k)$ are,
to good approximation,
independent of temperature at high frequencies,  
$| i\omega_n|\!\gg\!T$ and

(2) within that high-frequency regime, they are
slowly varying with $i\omega_n$, on $\omega_n$-scales
of order $T$ and that

(3) the contribution to $\Sigma(k)$ arising
from scattering into the high frequency region
[$|i\omega_{n^\prime}|\!\gg\!T$ in Eq.~(\ref{eqnself})],
denoted by $\Delta\Sigma(k)$ below, is to good approximation
independent of temperature and slowly varying for {\it all}
$i\omega_n$. 

For the case of the pure Hubbard model,
these NRG assumptions have been verified in great detail,
by explicit numerical calcuations \cite{pao94a}. 
A general justification of these assumptions
can be given. It is based on the notion that
the energy denominators in the Green's 
function $G(k)$ become very large and essentially 
$T$-independent for $| i\omega|\!\gg\!T$.
Hence, all strongly $T$-dependent details are 
``washed out'' in the high-frequency regime \cite{pao94a}.

As a consequence, only the low-frequency part of $\Sigma$, for
$| i\omega_n|\!\lesssim\!T$, arising from scattering into
the low-frequency region $|i\omega_{n^\prime}|\!\lesssim\!T$
in Eq.~(\ref{eqnself}), exhibits substantial 
$T$-dependence. As described in Ref.~\onlinecite{pao94a}, 
one therefore divides the self-energy 
$\Sigma(k)$ in Eq.~(\ref{eqnself})
into two contributions, arising respectively from 
the scattering $i\omega_n\!\to\!i\omega_{n^\prime}$
into a ``low'' region $L$
[i.e. $i\omega_{n^\prime}\, \in\, L$ in equation (\ref{eqnself})]
and into a ``high'' region 
[$i\omega_{n^\prime}\, \in\, H$  in equation (\ref{eqnself})], 
that is,
\begin{equation}
\Sigma (k)\ = \ {T\over N} 
\sum_{{\bf k^\prime}}
\sum_{i\omega_{n^\prime}\in L} S_\Sigma(k,k^\prime)
%
%
+\ \Delta \Sigma (k)
  \ .
\label{eqnselfrg}
\end{equation}
Here $S_\Sigma(k,k^\prime)$ denotes the
summand in Eq.~(\ref{eqnself}),
and $\Delta\Sigma$ is the contribution 
from the $i\omega_{n^\prime}$-summation over the 
``high'' region $H$. 

The basic idea of the NRG approach is
to reduce the numerical effort by evaluating 
$\Delta\Sigma(k)$ at a higher temperature, 
on a correspondingly coarser Matsubara grid. 
This higher-$T$ result is then interpolated
onto the finer Matsubara grid relevant to the lower $T$. 
The interpolation onto the lower-$T$
Matsubara grid needs to be performed only
for Matsubara frequencies $i\omega_n\!\in\!L$.
Only the ``low'' contribution in Eq.~(\ref{eqnselfrg}) 
needs to be evaluated by summing $i\omega_{n^\prime}$ over the
$L$-portion of the finer, lower-$T$ grid and this, again, needs to be
done only for $i\omega_n\!\in\!L$.

Starting from a large initial temperature $T_0$ and large initial
Matsubara summation cut-off $\Omega_0$,
this basic NRG step is carried out repeatedly,
through a sequence of decreasing temperatures 
\begin{equation} \label{eqnT_i}
T_0 > T_1 > ... > T_i > ...
\end{equation}
and decreasing Matsubara cut-offs 
\begin{equation} \label{eqnOmega_i}
\Omega_0 > \Omega_1 > ... > \Omega_i > ...\ \ ,
\end{equation}
until the desired final temperature is reached.
The initial maximal cutoff must be chosen large enough that
the physical results of the calculation, $\Sigma(k;T)$ and $\lambda(T)$,
are independent of $\Omega_0$, {\it i.e.}, typically large
compared to the bandwidth $8t$.

The subsequent renormalized $\Omega_i$ (with $i>0$) delineate the
boundaries between the low and high regions, $L_i$ and $H_i$,
in the $i$-th NRG step where
\begin{equation} \label{eqnL_i}
L_i \ =\ \{ i\omega^{(i)}_n \ {\bf |} \
\Omega_{i} > |\omega_n^{(i)}|; \ n\ {\rm integer} \}
\end{equation}
and
\begin{equation} \label{eqnH_i}
H_i\ =\ \bigcup_{j=1}^i \Delta H_j
\end{equation}
comprises the high-frequency region increments $\Delta H_j$
of the present ($j=i$) and all prior ($j<i$) NRG steps,
given by
\begin{equation} \label{eqnDH_j}
\Delta H_j\ =\ \{ i\omega^{(j-1)}_n \ {\bf |} \
\Omega_{j-1} > | \omega_n^{(j-1)}| > \Omega_j; \ n\ {\rm integer} \}
\end{equation}
The respective fermion Matsubara frequency grids are given by
\begin{equation}\label{eqnMeshes}
i\omega^{(j)}_n\ =\ (2n+1)\pi T_{j}\ .
\end{equation}
for integer $n$ and $j$.
In order to ensure that the summands in 
Eq.~(\ref{eqnDSigma}) below enter with the correct 
weights, the $\Omega_j$ and $T_j$ are chosen to 
obey the grid matching conditions
\begin{equation} \label{eqnGridMatch}
\Omega_j = 2 \pi N_j T_j,\ \ \ \ \
\Omega_j = 2 \pi K_j T_{j-1}
\end{equation}
so that
\begin{equation} \label{eqnTRat}
T_j/T_{j-1} = K_j/N_j, \ \ \ \ \ 
\Omega_j/\Omega_{j-1} = K_j/N_{j-1}
\end{equation}
with $N_j$ and $K_j$ integer and $K_j\ge N_j\ge 1$,
as described in detail in Ref.~\onlinecite{pao94a}.
The calculations presented below are based on a ``factor--2''
renormalization group with
$\Omega_j/\Omega_{j-1} = T_j/T_{j-1} =1/2$,
as illustrated by the frequency grids shown in 
Fig.~\ref{rgfreq}. Note that the resulting
NRG fermion Matsubara grid $L_i\cup H_i$ at temperature $T_i$ 
is substantially  ``diluted''  compared to the full, dense
Matsubara grid
\begin{equation} 
\label{eqnA_i}
A_i \ =\ \{ i\omega^{(i)}_n \ {\bf |} \
\Omega_0 > |\omega_n^{(i)}|; \ n\ {\rm integer} \} \ .
\end{equation}
The latter constitutes the basic frequency summation domain 
in a ``brute force'' calculation at temperature $T_i$.

The contribution from the $H_i$ region,
$\Delta \Sigma^{(i)}(k)$, is ``frozen in'' during the
self-consistency iteration at temperature $T_i$. It is calculated,
prior to the self-consistency iteration, by
\begin{eqnarray} 
\label{eqnDSigma}
\Delta \Sigma^{(i)} (k) \ &=&\ \sum^{i-1}_{j = 0}\,
{T_j\over T_i} \sum_{{\bf k^\prime}}
\sum_{i\omega_{n^\prime}^{(j)}\in \Delta H_{j+1}} 
S_\Sigma^{(j)} (k,k^\prime)
\\
\ &=&\ \Delta \Sigma^{(i-1)} + 
{T_{i-1}\over T_i} \sum_{{\bf k^\prime}}
\sum_{i\omega_{n^\prime}^{(i-1)}\in \Delta H_{i}} 
S_\Sigma^{(i-1)} (k,k^\prime)
\\
\label{eqnDSigma2}
\ &\equiv&\ \Delta \Sigma^{(i-1)} + \delta\Sigma^{(i)}
\ ,
\end{eqnarray}
where now $k'\equiv({\bf k'},i\omega_{n^\prime}^{(j)})$ and
$S_\Sigma^{(j)}(k,k^\prime)$ is the summand in 
Eq.~(\ref{eqnself}),
computed with the Green's function  $G^{(j)}(k^\prime)$,
which, in turn, is obtained via the Dyson equation (\ref{eqndyson})
from the self-energy $\Sigma^{(j)}(k^\prime)$ on the
grid $L_j\cup H_j$ at temperature $T_j$. Note here that the
linear interpolation of the summand from higher- to lower-$T$ 
frequency grids introduces the temperature 
re-weighting factor $T_j/T_i$
into the summation in Eq.~(\ref{eqnDSigma}).

The evaluation of the increment 
$\delta\Sigma^{(i)}$ via Eq.~(\ref{eqnDSigma2})
needs to be carried out only for $i\omega$-points in $L_{i-1}$. 
Both $\delta\Sigma^{(i)}$ and $\Sigma^{(i-1)}$ are then
added and interpolated onto the finer $i\omega_n^{(i)}$-grid,
inside $L_i$, to obtain $\Delta \Sigma^{(i)}$ on $L_i$. 
This fixed $\Delta\Sigma^{(i)}$ is then
used in the (re-)calculation of 
$\Sigma^{(i)}$ on $L_i$, via Eq.~(\ref{eqnselfrg}), 
during the self-consistency iteration at temperature $T_i$.

Note that $\Delta\Sigma^{(i)}({\bf k},i\omega_n^{(j)})$ 
does {\it not} have to be calculated or stored for 
grid points outside of the low-frequency domain $L_i$,
that is, for $i\omega_n^{(j)}\in H_i$.
The values of $\Sigma^{(i)}({\bf k},i\omega_n^{(j)})$
on $H_i$ are already ``frozen in'' before
the self-consistency iteration at temperatuure $T_i$, 
since, by the above-stated NRG assumption of $T$-independece
in $H_i$,
\begin{equation}
\label{eqnSigmaH}
\Sigma^{(i)}({\bf k},i\omega_n^{(j)})
\ =\ 
\Sigma^{(i-1)}({\bf k},i\omega_n^{(j)})
\ =\ ...
\ =\ 
\Sigma^{(j)}({\bf k},i\omega_n^{(j)})
\end{equation}
for $i\!>\!j$ and $i\omega_n^{(j)}\in H_i$. 
Hence, the values of $\Sigma^{(j)}$
on $H_j\subset H_i$ become ``frozen in'' 
({\it i.e.}, permanently stored)
after the $j$--th NRG step,
and need not be re-calculated
during subsequent NRG steps $i\!>\!j$. Only
the values of $\Sigma^{(i)}$ on the low-frequency grid
$L_i$ need to be re-calculated and iterated to self-consistency
during the $i$--th NRG step.

In order to evaluate the self-energy $\Sigma$,
we therefore need the effective potentials
$V_2$, $V^{ph}$ and $V^{pp}$ at $i\omega_n$ 
and $i\omega_{n^\prime}$ 
in the $L$--region only.
However, the fermion frequency summation 
over $i\omega_{n^{\prime\prime}}$ and the 
fermion frequency matrix inversions in equations 
(\ref{eqnvph}) and (\ref{eqnvpp}) still
run over the whole Matsubara frequency range up to the cutoff $\Omega_0$.
We therefore have to develop an efficient algorithm to overcome 
the fast growth of CPU time and memory requirements
associated with these matrix operations at low temperatures.
To this end, the NRG approach, 
outlined above for the self-energy $\Sigma(k)$,
must be extended to the fermion frequency matrices,
$R_{n,n^\prime}(q)$, where $R$ stands for $D$, $M$, $S$, or $T$,
as defined in Eq.~(\ref{eqnRmat}).
In the following, we will, for simplicity, omit the 
$q$-argument and use the notation
\begin{equation}
 R(i\omega_n,i\omega_{n^\prime})
 \equiv R(i\omega_n,i\omega_{n^\prime};q) \equiv R_{n,n^\prime}(q)\ .
\end{equation}

As defined in Eqs.~(\ref{eqnRmat}-\ref{eqnxpp}), 
these $R$-matrix elements can be regarded as the
values of an analytical function $R(i\omega,i\omega')$, defined
for continuous $i\omega$- and $i\omega'$-arguments.
The basic assumption underlying our NRG approach for the $R$-matrix 
is that ${1\over T} R(i\omega,i\omega')$, as defined in 
Eqs.~(\ref{eqnRmat})--(\ref{eqnxpp}) is 

(1) independent of temperature and 

(2) slowly varying on an $i\omega$-scale of order $T$,
\\
if {\it either} $| i\omega|\!\gg\! T$ {\it or} 
$| i\omega'|\!\gg\! T$ {\it or}  both. 
The justification for these assumptions lies again
in the high-frequency behavior of the energy denominators
of the Green's function $G(k)$, analogous to the NRG assumptions
for the self-energy. We can therefore use the same strategy
as in the NRG approach for the self-energy:

We again divide the full 
Matsubara frequency range into an $L$--
and an $H$--region and, in the $H$--region, we calculate the $R$-matrix
at a higher temperature on a correspondingly coarser grid.
For all required Matsubara frequency summations in the $H$-region, 
the $R$-matrix is then, again, interpolated onto the finer grid.
In detail, this works as follows:

The full $R$-matrix, denoted by $R^{(i)}$ at temperature $T_i$, 
is defined for matrix indices
$i\omega_n^{(i)}$ covering the full, dense $i\omega_n^{(i)}$-grid
$A_i$, Eq.~(\ref{eqnA_i}),
up to the maximum cutoff $\Omega_0$.
In all matrix multiplications during the $i$-th NRG step, 
the full $R^{(i)}$ is now replaced by a ``diluted'' $R$-matrix,
which needs to be evaluated and stored only for indices on
the NRG frequency grid $L_i\cup H_i$.
Those matrix elements which are eliminated by this procedure
from the full $R$-matrix are approximated by appropriate
inter- and extrapolations from the $R^{(i)}$-matrix elements retained.

Consider, for example, a typical matrix-vector multiplication
of the $R$-matrix with a fermion frequency vector $f(i\omega)$.
At the $i$--th NRG step, the required
summation over all Matsubara frequency matrix indices $i\omega_n^{(i)}$
up to the cutoff, $|\omega_n^{(i)}|<\Omega_0$, is replaced
by summations over the ``diluted'' frequency grid $L_i\cup H_i$. 
That is, rather than carrying out the matrix-vector multiplication on 
the full, dense $A_i$-grid to obtain
\begin{equation} 
\label{eqnRmult}
g(i\omega_n^{(i)})
\ \equiv\ \sum_{i\omega_{n^\prime}^{(i)}\in A_i}
R^{(i)}(i\omega_n^{(i)},i\omega_{n^\prime}^{(i)})
\ f(i\omega_{n^\prime}^{(i)})\ ,
\end{equation}
we evaluate instead $g(i\omega_n^{(j)})$ on the NRG grid
$i\omega_n^{(j)}\in L_i\cup H_i$
by linear interpolation of the summand in the $H_i$--region,
which yields, similar to Eq.~(\ref{eqnDSigma}),
\begin{eqnarray} 
\label{eqnRmultRG}
g(i\omega_n^{(j)})
\ &=&
\ \ \ \ \sum_{i\omega_{n^\prime}^{(i)}\in L_i}
R^{(i)}(i\omega_n^{(j)},i\omega_{n^\prime}^{(i)}) 
\ f(i\omega_{n^\prime}^{(i)})
\nonumber
\\
&\ &\ \ \ +\ 
\sum_{j^\prime=0}^{i-1}
\sum_{i\omega_{n^\prime}^{(j^\prime)}\in \Delta H_{j^\prime+1}}
{T_{j^\prime}\over T_i}
R^{(i)}(i\omega_n^{(j)},i\omega_{n^\prime}^{(j^\prime)}) 
\ f(i\omega_{n^\prime}^{(j^\prime)})
\nonumber
\\
&=&
\ \sum_{i\omega_{n^\prime}^{(j^\prime)}\in L_i\cup H_i}
{T_{j^\prime}\over T_i}
R^{(i)}(i\omega_n^{(j)},i\omega_{n^\prime}^{(j^\prime)}) 
\ f(i\omega_{n^\prime}^{(j^\prime)}) \ .
\end{eqnarray}
Note that the interpolation again introduces a temperature
re-weighting factor $T_{j^\prime}/T_i$ into the summation
over the coarsened grid.
The values of $g(i\omega_n^{(i)})$ and $f(i\omega_n^{(i)})$
on the original full $A_i$--grid
are then again representable by interpolation in terms of
the $g(i\omega_n^{(j)})$- and $f(i\omega_n^{(j)})$-values,
respectively, on the NRG grid $L_i\cup H_i$.

This NRG grid representation of the $R$-matrix,
Eq.~(\ref{eqnRmultRG}),
is also used to carry out the matrix inversion of $Q\!\equiv\!1+R$
entering into Eqs.~(\ref{eqnvph}) and (\ref{eqnvpp}).
From Eq.~(\ref{eqnRmultRG}) it is easy to see that,
at the $i$--th NRG step, the problem is reduced to 
carrying out the matrix inversion of a ``diluted'' 
$Q$-matrix with matrix elements
\begin{equation}
\label{eqnQmat}
Q^{(i)}(i\omega_n^{(j)},i\omega_{n^\prime}^{(j^\prime)})
=\delta_{n,n^\prime}\delta_{j,j^\prime} +
{T_{j^\prime}\over T_i}
R^{(i)}(i\omega_n^{(j)},i\omega_{n^\prime}^{(j^\prime)})\ ,
\end{equation}
where the matrix indices 
$i\omega_n^{(j)},i\omega_{n^\prime}^{(j^\prime)}$
are restricted to the NRG grid $L_i\cup H_i$.

Based on the above-stated NRG assumption of approximate 
$T$-independence of $R/T$ in the high-frequency region, 
we can express the $R^{(i)}$ matrix elements 
in the ``$H_i$--$H_i$'' region by $R$-matrix 
elements already calculated in previous NRG steps. 
That is, analogous to Eq.~(\ref{eqnSigmaH}), we have
\begin{equation}
\label{eqnRmatH}
R^{(i)}(i\omega_n^{(j)},i\omega_{n^\prime}^{(j^\prime)}) 
\ =\ 
{T_i\over T_{i-1}}
R^{(i-1)}(i\omega_n^{(j)},i\omega_{n^\prime}^{(j^\prime)})
\ =\ ...
\ =\ 
{T_i\over T_{j^{\prime\prime}}}
R^{(j^{\prime\prime})}(i\omega_n^{(j)},i\omega_{n^\prime}^{(j^\prime)})
\end{equation}
if both $i\omega_n^{(j)}$ and $i\omega_{n^\prime}^{(j^\prime)} \in H_i$.
Here $j^{\prime\prime}$ denotes the larger of $j$ and $j^\prime$.
In other words, after the $j$-th NRG step, the $H_j$--$H_j$ matrix 
elements of $R^{(j)}$ are ``frozen in'' and need not be
recalculated in subsequent NRG steps $i>j$, except for a 
change in the temperature prefactor. Only those matrix elements
of $R^{(i)}$ connecting $L_i$ to $H_i$ and $L_i$ to $L_i$ 
need to be calculated and updated to self-consistency during 
the self-consistency iteration of the $i$--th NRG step.

In the pairing eigenvalue calculation, 
Eq.~(\ref{eqnlambda}), we can use the same interpolative
approach described above for the $R$-matrix,
to carry out matrix multiplications with the pairing
kernel, exploiting the diluted, but non-equidistant
NRG frequency grid $L_i\cup H_i$ at the final temperature
$T\!=\!T_i$. The non-equidistant nature of the grid
will again introduce temperature re-weighting factors
into the Matsubara summation, analogous to 
Eq.~(\ref{eqnRmultRG}).

The dimension of $Q^{(i)}$ on the NRG grid $L_i\cup H_i$ is
${\cal N}(L_i\cup H_i)\!\times\!{\cal N}(L_i\cup H_i)$
where ${\cal N}(L_i\cup H_i)$ denotes the total number of
grid points in $L_i\cup H_i$. The inversion of $Q^{(i)}$
at all momentum-energy-transfer vectors $q$
[see Eqs.~(\ref{eqnvph}--\ref{eqnRmat})] is by far the most CPU
time consumptive step at low temperatures.
In, say, a ``factor--2'' NRG calculation,
${\cal N}(L_i\cup H_i)$ at low temperatures
increases with $1/T$ as ${\rm log}_2({1\over\pi}\Omega_0/T)$;
the amount of CPU
time for inversion of a general $D\!\times\!D$ matrix 
scales as $D^3$; 
there are of order 
$N_\omega\!\times\!N\!\times\!N_{\rm it}$ 
such matrix inversions [one per $q$-vector and self-consistency
iteration] in each NRG step,
where $N$ is the spatial lattice ($\equiv$ ${\bf k}$ grid) size,
$N_\omega=\Omega_0/(\pi T)$ is the size of the original 
Matsubara frequency grid for maximal cutoff $\Omega_0$
and final temperature $T$,
and $N_{\rm it}$ is a typical number of self-consistency
iterations needed per NRG step;
and the number of NRG steps to reach the final temperature
$T$ scales as ${\rm log}_2({1\over\pi}\Omega_0/T)$.
The total CPU time of the full NRG low-$T$ self-energy calculation 
therefore scales as
$N_{\rm it}\!\times\!N\!\times\!({1\over\pi}\Omega_0/T)\!\times\! 
[{\rm log}_2({1\over\pi}\Omega_0/T)]^4$.
This should be compared to the CPU time of a brute force
calculation which, estimated along similar lines, scales 
at least as 
$N_{\rm it}\!\times\!N\!\times\!({1\over\pi}\Omega_0/T)^4$.
The savings in CPU time of the NRG approach relative to a
brute force calculation is therefore a factor of order
$({1\over\pi}\Omega_0/T)^3/[{\rm log}_2({1\over\pi}\Omega_0/T)]^4$.

The memory requirements for both NRG and brute force method
are dominated by the storage of the $R^{(i)}$- and $Q^{(i)}$-matrices.
Following the foregoing dimensionality estimates, this scales as 
$N\!\times\!({1\over\pi}\Omega_0/T)\!\times\!
[{\rm log}_2({1\over\pi}\Omega_0/T)]^2$ 
in the NRG and as
$N\!\times\!({1\over\pi}\Omega_0/T)^3$
in the brute force approach. 
Thus, in terms of memory
consumption, the NRG saves a factor of order
$({1\over\pi}\Omega_0/T)^2/
[{\rm log}_2({1\over\pi}\Omega_0/T)]^2$ 
relative to the brute force approach.

\section{Results}
\label{sec:result}

To begin with, we explore the frequency 
and temperature dependence of the 
self-energy and Green's function of the pure Holstein model
($U=0$) in Fig. \ref{self1}, using a brute force
calculation for temperatures 
$T/t\,=\, 1.0, \, 0.5,\, 0.25,\, {\rm and}\, 0.125$.  
We use a maximum cutoff frequency
$\Omega_0/t\, =\, 25$, 
a $16^2$ momentum grid, and an
electron concentration $\langle n\rangle = 1.0$. 
In Fig. \ref{self1}(a) and (b), we plot the real part and
imaginary part of the self-energy, respectively, as a function 
of Matsubara frequency 
at ${\bf k}~=~(1.77,1.77)$, which is a point just above the 
Fermi surface along the diagonal direction.
As expected in the high frequency region, the temperature
dependence of the self-energy is much smaller than in 
the low frequency region. Note that we plot the imaginary part of 
as $\Sigma_2 / \omega_n$ instead of $\Sigma_2 / t$, which emphasizes 
the low-$i\omega$ region at low temperature.
 
Fig. \ref{self1}(c) shows the Green's 
function at these parameters. 
It is quite clear that the high temperature 
behavior of the one--particle function is ``frozen in'' 
very quickly as the temperature decreases. 
As explained in the previous
section, this is very important for the fermion frequency matrix 
NRG method to be applicable, since fluctuation propagators 
in the effective or pairing potentials, Eqs.~(\ref{eqnv2}-\ref{eqnvpp})
and (\ref{eqnvpair}),
are calculated from the one--particle Green's function. 
Clearly, at high frequencies the Green's function can be interpolated
by the high temperature results without loosing any informations.

At Fig. \ref{self3}, we plot the self-energy and Green's function 
right at the Fermi surface for 
${\bf k}\, =\, (2.95,0.20)$. At half--filling, the real parts of 
self-energy and of Green's function vanish at the Fermi surface,
due to particle-hole symmetry.
The low frequency behavior becomes much sharper at the Fermi surface, 
but the high frequency parts of $\Sigma_2$ and $G_2$ are almost temperature
independent. 

In order to test the accuracy of our NRG approach, 
we use the factor--2
frequency NRG operation with 
constant $N_j=4$ and $K_j=8$,
starting at $T_0/t\,=\,1.0$, as described 
in the previous section.
After three such
NRG operation, we thus reach the final $T\equiv T_3\, =\, 0.125t$.
We plot the self-energy and
Green's function for two different momentum points 
in Fig. \ref{self1rg}, for ${\bf k} = (1.77,1.77)$, and  
in Fig. \ref{self3rg}, for ${\bf k} = (2.95,0.20)$. 
For comparison, we show the brute force results obtained
for the same model parameters, temperature $T$, and 
Matsubara cutoff $\Omega_0\, =\, 25.13$ (fixed).
There is remarkable  agreement between the two methods.
However, the NRG grid $L_i$, Eq.~(\ref{eqnL_i}),
to be summed over in the self-consistency iterations,
contains only $2N_i=8$ fermion Matsubara frequencies
at each NRG step $i$,
down to $T=T_3=0.125$ and $\Omega_3\, =\, 3.1$. 
By contrast, in the brute force approach one has 
to sum over the full, dense $A_i$-grid, Eq.~(\ref{eqnA_i}),
with $\Omega_0/(\pi T)=64$ 
fermion Matsubara frequencies. 
In table \ref{rg2bruteforce}, we list the memory and
CPU time per iteration for both approaches.
Savings of memory and CPU time of about two orders of magnitude
are achieved by the NRG approach, without significant loss
in accuracy. 

Next, we examine the stability of the NRG results
against changes in the NRG control parameters and NRG protocol.
Fig. \ref{eigen2t} shows the $s$--wave
eigenvalue calculation for three different ``lower cutoff''
frequencies $\Omega_\ell/t\, =\,$ 12.57, 6.283, and 3.142,
which correspond to carrying out a total of, respectively, 
$\ell=3$, 4, and 5 factor--2 NRG steps, each starting from 
$T_0/t\!=\!4.0$ and $\Omega_0/t\!=\!100.5$, with fixed
$N_0=N_1=...=N_\ell=4$. In each of these three calculations,
the $\ell$--th factor--2 NRG step is followed by one
``fixed-cutoff'' step where the cutoff, $\Omega_{\ell+1}$,
is left unchanged at $\Omega_{\ell+1}=\Omega_\ell$,
while $N_{\ell+1}$ is increased, beyond $N_\ell$, 
in order to lower the final temperature 
$T=T_{\ell+1}$.\cite{pao94a}
This last, $(\ell+1)$--th step is repeated with 
several different values of $N_{\ell+1}$, in order to scan
the low-$T$ regime. 

The $s$-wave $T_c$ is then determined by interpolation 
between two adjacent low-$T$ points, $T_+$ and $T_-$, say,
which bracket the instability, that is, 
their pairing eigenvalues,
$\lambda(T_+)<1<\lambda(T_-)$, bracket unity. 
The $s$--wave transition temperatures estimated from the three
different calculations are $T_c/t=0.093$, 0.090, and 0.091
for $\ell=3$, 4, and 5, respectively.
Thus, both the $T_c$ and the $\lambda(T)$ results 
[Fig.~\ref{eigen2t}] 
obtained with the three different lower cutoff protocols, 
$\ell=3$, 4, 5, are in excellent ($\sim 1-2\%$) agreement.

For all further $T_c$ results discussed below, the same
``factor--2 plus fixed-$\Omega_\ell$'' NRG protocol is used,
with an initial temperature $T_0/t \, =\, 4.0$,
initial cutoff $\Omega_0/t \, =\, 100.5$, and
$\ell=5$ of factor--2 NRG steps, corresponding to 
$T_\ell/t\, =\, 0.125$ and a lower cutoff 
$\Omega_\ell/t\, =\, 3.142$.

Fig. \ref{tc2up} shows the $s$--wave $T_c$ as a function of the 
EP coupling strength $U_p$. A 16$^2$ grid is employed.
The Einstein phonon frequency is $\Omega_p=1.0t$. 
Within the FLEX approximation, the $s$--wave $T_c$ of the Holstein
model increases with $U_p$ over a wide range of the EP coupling and
saturates when the $U_p$ is comparable to the band width (8t).

The finite size effect at the low temperatures needs to be 
treated carefully.
We therefore calculate the $s$--wave transition 
temperature as a function of electron concentration 
$\langle n\rangle$ in a wide filling range 
near half--filling. 
The results are shown in Fig. \ref{stc2n} for
three different ${\bf k}$--grids, $N=16^2$, 32$^2$, and $64^2$.
The 16$^2$ grid gives reasonable estimates over
for a wide temperature range, down to the lowest temperatures
we reach in the calculation, $T\sim0.02t$. 
A 32$^2$ grid, in general, covers this whole temperature range,
down to $0.02t$ very accurately.
 
Three sets of $s$--wave $T_c$ are reported here corresponding to 
an on--site Coulomb $U/t\,=\, 0$ (Holstein model), 2, and 3. 
All of them have the same Einstein phonon frequency
$\Omega_p/t\, =\, 1.0$ and EP coupling $U_p/t\, =\, 4.0$.
The presence of the on--site Coulomb 
repulsion suppresses the $s$--wave 
pairing and $T_c$ goes to zero or becomes
smaller than our lowest numerically accessible $T$
($\sim0.02t$), when $U$ approaches $U_p$.
Thus, in essence, $T_c$ goes to zero (or a numerically ``very small''
value) when the on-site Coulomb repulsion $U$ overcomes the
phonon-mediated on-site attraction $U_p$.
Note that there is no significant reduction
of the $s$-wave-supressing $U$ effect due to 
retardation, that is, due to the 
``pseudo-potential'' reduction of the
Coulomb repulsion.\cite{scal69,allen_mitr}
This is perhaps not surprising, since, on the one-hand,
the phonon-frequency is quite sizeable here, 
compared to the band-width $8t$, and, on the other hand,
near $1/2$-filling there may be relevant electronic 
(charge, spin, and/or pair) flucutation energy scales 
in the problem which are even closer to the phonon
energy scale.
Note also that, there exists an ``optimal'' doping, of
about 20\% to 30\%, where
a maximum $s$--wave transition temperature $T_c$ 
occurs for this model in the FLEX approximation. 

A comparison between FLEX and conventional
Eliashberg theory in the $s$--wave pairing regime
$U<U_p$, is shown in Fig. \ref{stc2n1}.
Here, we have used the bare potential 
$U+v_p(i\omega_n-i\omega_{n^\prime})$
as the effective exchange potential in the
self-consistent (Migdal) self-energy calculation
and as the pairing potential $V_{pair}$
in the Eliashberg pairing eigenvalue calculations.
Thus, in essence, our Eliashberg calculation
negelects all the screening effects due to electronic
particle-hole and particle-particle fluctuations
which the FLEX approximation seeks to include.
At sufficiently large doping, $\gtrsim 15-20\%$,
where converged results for $T_c$ can be obtained in both
approaches, the FLEX $T_c$ is noticeably higher
that the Eliashberg $T_c$. This suggests that
the predominant ({\it i.e.}, charge) fluctuations
included in FLEX enhance the $s$-wave pairing potential. 
Because of the lack of screening in the Eliashberg
(and because of its presence in the FLEX) calculation,
it is not surprising that the relative discrepancy between
the two approaches becomes even larger in the presence
of a finite on--site $U=2.0t$, as shown by the dashed line results
in Fig. \ref{stc2n1}.

A physically very interesting problem to study 
in the Holstein--Hubbard model is 
the competition between $d$-- and $s$--wave singlet pairing.
While the $d_{x^2-y^2}$ pairing 
instability is by now well-established in several
high--$T_c$ cuprates, there is also mounting 
(although by no means unambiguous) experimental evidence 
for $s$--wave pairing in some of these materials.\cite{scal95,legg96} 
On the theoretical side, 
it is, within a weak-coupling self-consistent diagrammatic
framework, well-established that a
$d_{x^2-y^2}$ pairing instability can be driven, or at least enhanced
by AF SF exchange,\cite{bick87,bick89,bick91,yone89,moriya,pines} 
while being suppressed, due to self-energy
effects, by the presence of phonon exchange\cite{schu95,pao_tbp}.
An $s$--wave instability, on the other hand, can be driven 
or enhanced by phonon exchange, while being suppressed
by AF SF exchange and local Coulomb repulsions, as already 
discussed above. 
Thus the two possible candidate pairing mechanism
for $d$-- and $s$--wave pairing tend to be mutually destructive.

The Holstein--Hubbard model, treated in the FLEX approximation, 
may be a reasonable starting point to investigate the magnitude
these mutually destructive ``anti-pairing'' effects, both on the 
$s$--wave and on the $d$--wave side of the phase boundary.
In Fig. \ref{stc2n}(a) and (b), we plot the $s$-- and, respectively,
$d$--wave phase diagram at a fixed $U_p=4.0t$ and, respectively, 
fixed $U=4.0t$, for on--site Coulomb repulsions $U/t=0$, 2 and 3,
and respectively, for on--site attraction $U_p/t=0$, 2 and 3. 
The Einstein phonon frequency is fixed at $\Omega_p/t\, =\, 0.5$
for both cases. 
In comparing the two pairing states, we note that the
$s$--wave state which has the higher $T_c$
in the absence of its ``anti-pairing'' interactions
($U=0$) is also suppressed more strongly, when its anti-pairing
interaction $U$ is turned on. 
We note also that the $s$--wave instability exists over a wider range 
of doping $1-\langle n\rangle$. An optimal doping 
with very broad $T_c$ maxima is found in both
$s$-- and $d$--wave pairing states.
In the $s$--wave case this occurs at 20\% to 30\% hole dopings, 
depending on the on--site Coulomb repulsion $U/t$;
in the $d$--wave case the $T_c$-maximum is near $10\%$ doping with
a much less pronounced dependence on $U_p$.

Another important feature to study in the Holstein--Hubbard model 
is the isotope effect, that is, the dependence of 
the superconducting $T_c$ on the isotopic mass $M$ of the ions.
In our parametrization of the model, this isotopic mass dependence
enters only via the Einstein phonon frequency $\Omega_p$,
that is, via Eq.~(\ref{eqnOmega_P}), since all other
parameters ($t$, $U$, $C$, $K$, $U_p$) are of purely
electronic origin, {\it i.e.}, not dependent on $M$.
An important experimental measure of the
lattice effects on $T_c$ is the isotope exponent
\begin{equation}
\label{eqnAlpha}
\alpha\ =\ -{ \partial \ln T_c \over \partial \ln M}\Bigg|_e
\ =\ { 1 \over 2} {\partial \ln T_c
  \over \partial \ln \Omega_p}\Bigg|_e \ ,
\end{equation}
where the notation $...\big|_e$ means that 
the partial derivative should be taken with all 
above-identified electronic parameters held constant.
The second equality in (\ref{eqnAlpha}) follows
from Eq.~(\ref{eqnOmega_P}).

In Fig. \ref{stc2w0}, we plot both the  $s$-- and $d$--wave
superconducting transition temperatures as a function
of the phonon frequency $\Omega_p$. In the $s$--wave
case, Fig.~\ref{stc2w0}(a), 
$T_c$ rises approximately linearly with $\Omega_p$,
up to $\Omega_p$ of about $2-3t$. At larger $\Omega_p$,
$T_c$ gradually becomes sub-linear
and approaches saturation 
which is reached when $\Omega_p$ becomes of the order of
the electronic bandwidth, that is, in physical terms,
unrealistically large. The linear $\Omega_p$-dependence
of the $s$--wave $T_c$ implies that the isotope exponent
is given essentially by its classical BCS value
for conventional phonon-mediated $s$-wave superconductivity,
\begin{equation}
\alpha\ \cong\ {1\over2}\ ,
\end{equation}
in the physically relevant low-$\Omega_p$
regime $\Omega_p\lesssim t/4$.

In the $d$--wave case, $T_c$ has only a very weak,
slightly decreasing $\Omega_p$ dependence. 
The $d$--wave isotope exponent is therefore negative
and and small in magnitude, typically with
\begin{equation}
|\alpha|\ <\ 0.05\
\end{equation}
in the physically realistic $\Omega_p$-regime
$\Omega_p\lesssim t/4$.
This is much smaller than typical orders of magnitudes
$|\alpha|\sim0.4-1.0$  observed in non-optimally
doped cuprates, but confirms the conclusions from earlier 
calculations of the isotope exponent due to
harmonic phonon exchange in diagrammatic
$d$--wave pairing models.\cite{schu95,pao98}
\section{summary}
\label{sec:summary}

In summary, we have developed an important generalization
of the numerical renormalization
group (NRG) technique for solving 
the self-consistent field equations 
of the fluctuation exchange (FLEX) approximation
in the presence of a phonon-mediated, retarded bare interaction
potential. In the presence of retarded bare interactions
neither fast Fourier transform\cite{sere91}
nor the previously developed NRG approach\cite{pao94,pao94a} 
for purely instantaneaous interactions can be employed.
On the other hand, our generalized ``fermion 
frequency matrix'' NRG technique,
produces large gains in computational efficiency,
both in terms of CPU time and in terms of memory requirements,
relative to a brute force calculational approach.

In the physically most interesting low temperature regime,
the CPU time and memory requirements of the brute force approach 
exceed by far the limits of currently available computational resources.
By contrast, our generalized NRG method yield efficient, accurate
solutions and allows detailed studies of superconducting
instabilities in this regime, down to temperature 
scales 3 orders of magnitude below the electronic bandwidth.
Our  work also suggests possible avenues towards solving more complicated 
self-consistent approximation schemes, such the parquet 
approximation \cite{bick91,domi64}.

We have tested and applied this NRG approach in the context of the 
FLEX approximation to the the 2D Holstein--Hubbard model.
In this model, the FLEX equations are simplified
due to the lack of momentum (${\bf q}$-) dependence in the bare
Coulomb and in the bare phonon-mediated interaction potentials.
However, more general electron-phonon interaction models, including
${\bf q}$-dependent potentials can easily
be accommodated in our NRG approach, by combining it
with computationally efficient real-space
basis representations\cite{esir97,esir98,hank74}.
Such real-space basis representations have recently
been implemented with great success to solve the
FLEX equations in extended Hubbard model systems
with instantaneous, but ${\bf q}$-dependent
bare interaction potentials.

By varying the on--site Coulomb repulsion and EP coupling strength
we have studied the competing $s$-- and $d$--wave 
superconducting instabilities of the 2D Holstein--Hubbard 
model in the FLEX approximation.
An optimal $T_c$ for both cases was observed at about 20\% to 30\% doping
for $s$--wave pairing and 10\% for $d$--wave pairing.
The $s$--wave phase is favored when the phonon-mediated
on-site attraction $U_p$ exceeds the on--site Coulomb repulsion $U$
and its transition temperature is suppressed by increasing $U$.
Likewise, the $d$--wave phase is favored when 
the on--site Coulomb repulsion $U$ exceeds
the phonon-mediated on-site attraction $U_p$
and its transition temperature is suppressed by increasing $U_p$.
When $U\, \sim \, U_p$, the $T_c$'s of both pairing states
are suppressed to zero or to a numerically inaccesible very-low
temperature regime.

Finally, the isotope 
exponent $\alpha$ for the $s$--wave state is BCS like,
that is, $\alpha\cong {1\over 2}$,
at realistic phonon frequencies 
$\Omega_p/t\, \lesssim\, 0.25$.
In the $d$--wave state, the isotope exponents
are negative and small in magnitude, with typically
$|\alpha|<0.05$ in the physical phonon frequency regime
$\Omega_p/t\, \lesssim\, 0.25$.
The overall magnitude of $\alpha$ is far too small to
explain observed isotope data in non-optimally
doped cuprates. Our full FLEX 
results thus support the conclusions of earlier 
$d$--wave isotope 
calculations by the present authors.\cite{schu95,pao_tbp,pao98}

We would like to thank Prof. N.E. Bickers for many
helpful discussions.
This work was supported by the
National Science Council (NSC, Taiwan, R.O.C.) under Grant Nos.
872112--M194011 and  882112--M194001 (C.--H. P.), by the U. S.
National Science Foundation under Grant No.
DMR--9215123 (H.--B. S.) and by 
computing support from UCNS, University of Georgia,
and National Center for High--Performance Computing (NCHC, Taiwan, R.O.C.)
are gratefully acknowledged.




\begin{figure}
\caption{(a) The bare interactions in the Holstein--Hubbard model, which include
the on--site Coulomb repulsion $U$ and the EP coupling $v_{p}(i \nu_m)$. 
(b)--(d) show the bare vertices
of the Holstein--Hubbard model in FLEX approximation. These diagrams include 
contributions
(b) $V^D$ from density, and (c) $V^{M}$ from magnetic 
particle--hole fluctuations as well as
(d) $V^S$ from singlet and $V^T$ from triplet 
particle--particle fluctuations.
In (d), the upper ($+$) sign pertains to $V^S$,
the lower ($-$) sign to $V^T$. }  
 \label{vert}
\end{figure}

\begin{figure}
\caption{Imaginary frequency discretization for the frequency NRG space.
(a) Initial stage of the frequency space with cutoff $\Omega_0$, including
four positive and four negative fermion frequencies, corresponding
to $N_0=4$. (b) After one 
``factor--2'' NRG operation, the lower cutoff is 
$\Omega_1\,=\, \Omega_0/2$. There
are eight frequencies in ``$L_1$'' and four frequencies in ``$H_1$''.
(c) After two ``factor--2'' NRG operations, the lower cutoff becomes
$\Omega_2\, =\, \Omega_0/4$. There are 8 frequencies in ``$L_2$'' and
eight frequencies in ``$H_2$''.}
\label{rgfreq}
\end{figure}

\begin{figure}
\caption{Imaginary part of 
self-energy $\Sigma(k)\, =\, \Sigma_1(k) + i \Sigma_2(k)$ 
and Green's function $G(k)$ for a brute force calculation at four different
temperatures: $T/t\, =\,$ 1.0 (cross symbols), 0.5 (squares), 
0.25 (circles), and 0.125 (lines) at {\bf k}~=~(1.77,1.77). 
(a) Real part of self-energy
$\Sigma_1$. (b) Imaginary part of self-energy $\Sigma_2/\omega_n$ (c) The real 
part (solid symbols and line) and imaginary part
(open symbols and dashed line) of $G$. The parameters are: 
$\Omega_0\, =\, 25t$, $U/t\, =\, 0$, and $\langle n\rangle\, =\, 1.0$.  } 
\label{self1}
\end{figure}

\begin{figure}
\caption{The same parameters as in Fig. 3 with the {\bf k}~=~(2.95,0.20),
which is right at the Fermi surface. Plots are shown for imaginary parts of
self-energy and Green's function only; the real parts of self-energy
and Green's function vanish on the Fermi surface 
at $\langle n\rangle\, =\, 1.0$, due to particle-hole symmetry.} 
\label{self3}
\end{figure}

\begin{figure}
\caption{Comparison of the self-energy [(a) for real part and (b) for 
imaginary part] and Green's
function [(c) solid line for real part and dashed lines for imaginary part] 
using factor--2 frequency NRG and a
brute force approach. Results from 3 stages of NRG are represented
by symbols. Results from brute force are represented by lines. 
The parameters are the same as Fig. 3.}
\label{self1rg}
\end{figure}

\begin{figure}
\caption{Comparison of (a) the self-energy and (b) Green's
function 
using factor--2 frequency NRG and a
brute force approach. Results from 3 stages of NRG are represented
by symbols. Results from brute force are represented by lines. 
The parameters are the same as Fig. 4}
\label{self3rg}
\end{figure}

\begin{figure}
\caption{Maximal eigenvalue $\lambda_s$ 
of the pairing kernel Eq.~(\ref{eqnlambda})
in the $s$--wave symmetry channel for
three different lower cutoffs: 
$\Omega_\ell/t\, =\,$12.56 (solid line);
6.283 (cross symbols), and 3.142 (open circles). 
The model parameters are:
$\Omega_p/t\, =\, 5.0$, $U_p/t\, =\, 4.0$, $U/t\, =\, 2.0$,
and $\langle n\rangle\, =\, 0.75$} 
\label{eigen2t}
\end{figure}

\begin{figure}
\caption{The $s$--wave transition temperature $T_c$
as a function of the 
EP coupling strength $U_p$ for the Holstein model ($U\, =\, 0$)
with Einstein phonon frequency $\Omega_p/t\, =\, 1.0$ and electron
filling factor $\langle n\rangle\, =\, 0.75$.} 
\label{tc2up}
\end{figure}

\begin{figure}
\caption{Phase diagrams for the $s$--wave 
superconductivity of the Holstein--Hubbard
model for different on--site Coulomb repulsion. 
The model parameters are $U_p/t\, =\, 4.0$ and
$\Omega_p/t\, =\, 1.0$. The on--site Coulomb repulsion $U/t\, =\,$
0, 2.0, 3.0 (from the top curve to the bottom one). Three different
${\bf k}$--grid sizes are employed.}

\label{stc2n}
\end{figure}

\begin{figure}
\caption{Comparison of Eliashberg (open squares)
and FLEX (open circles) solution for the phase diagrams for 
the $s$--wave superconductivity of Holstein--Hubbard model. The parameters
are the same as Fig. 9 but data are shown only for 
$U/t\, =\, 0$ (solid lines) and 2 (dashed lines).
  }
\label{stc2n1}
\end{figure}

\begin{figure}
\caption{Superconducting phase diagram of the
Holstein--Hubbard model for 
(a) s--wave pairing with $U_p=4.0t$ and several different $U$
(b) $d$--wave with $U=4.0t$ and several different $U_p$.}
\label{stc2n2}
\end{figure}

\begin{figure}
\caption{(a) $s$--wave and (b) $d$--wave transition temperatures $T_c$
as functions of the Einstein phonon frequency $\Omega_p$ near the 
optimal doping ($\langle n\rangle\, =\, $ 0.75 for $s$--wave 
and 0.90 for $d$--wave).
In (a), $U_p/t=4.0$ and $U/t=0$ (solid line) and $U/t=2.0$ (dashed line).
In (b), $U/t=4.0$ and $U_p/t=2.0$.}
\label{stc2w0}
\end{figure}



\begin{table} 
\vspace*{0.5in}
\caption{
Comparison of memory  and CPU time requirements
between brute force and frequency NRG calculations. 
Parameters are the same as in Fig. 5.
The calculations were performed on an IBM RS6000/397 workstation.}
\vspace{0.5in}
\begin{tabular}{cdd}
    & memory & CPU time per iteration \\
\hline
 brute force & 105 MB & 78.1 sec\\
 frequency NRG & 9 MB  & 0.6 sec
\end{tabular}
\label{rg2bruteforce}
\end{table}

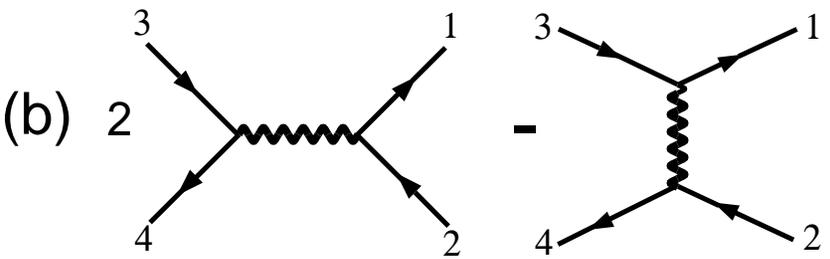
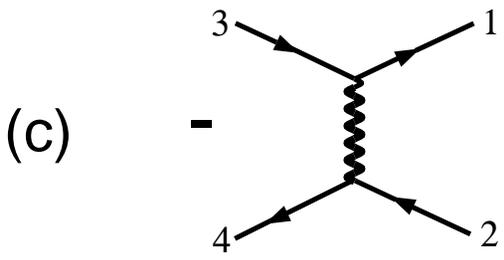
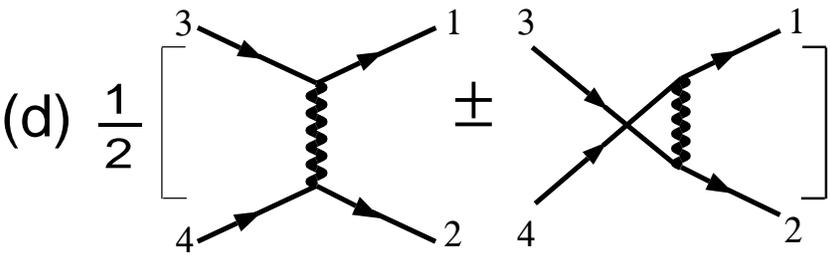
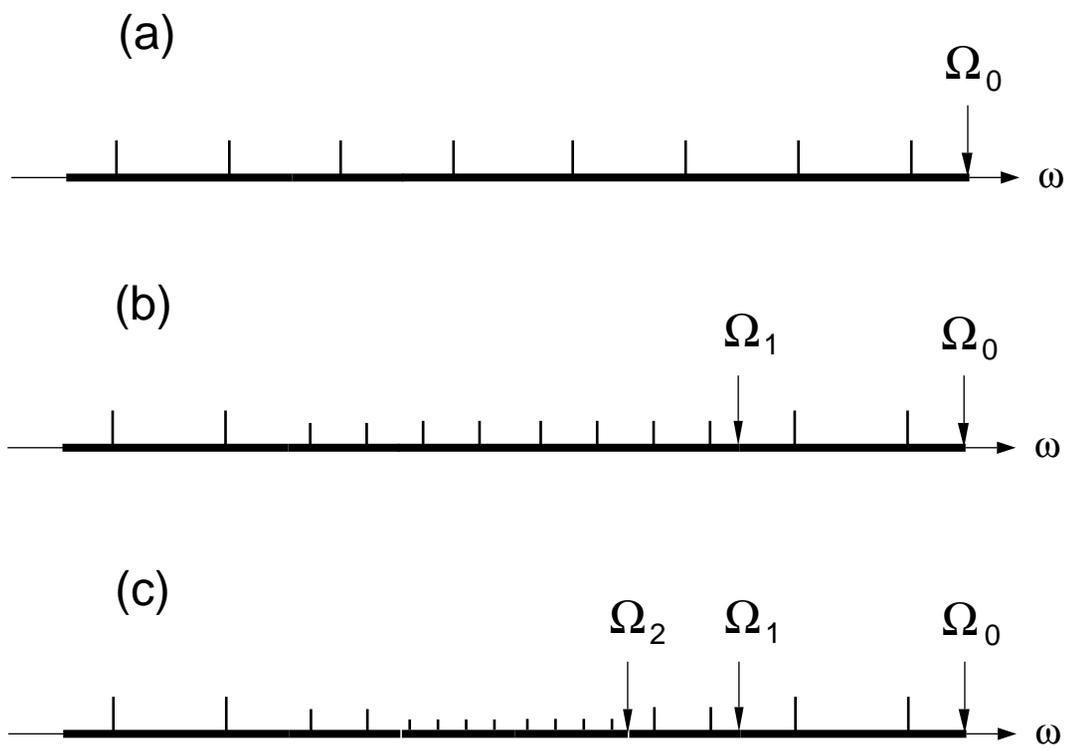
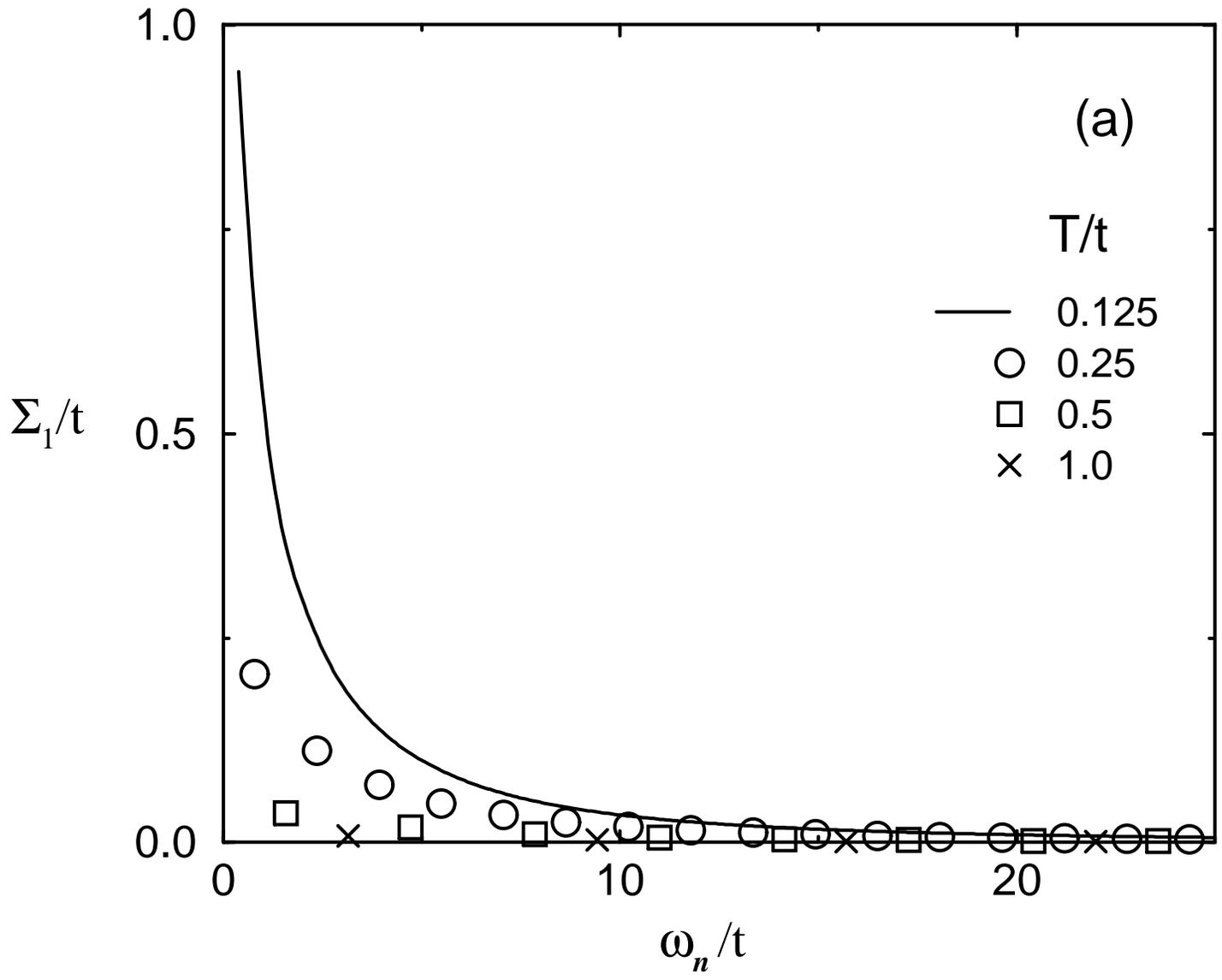
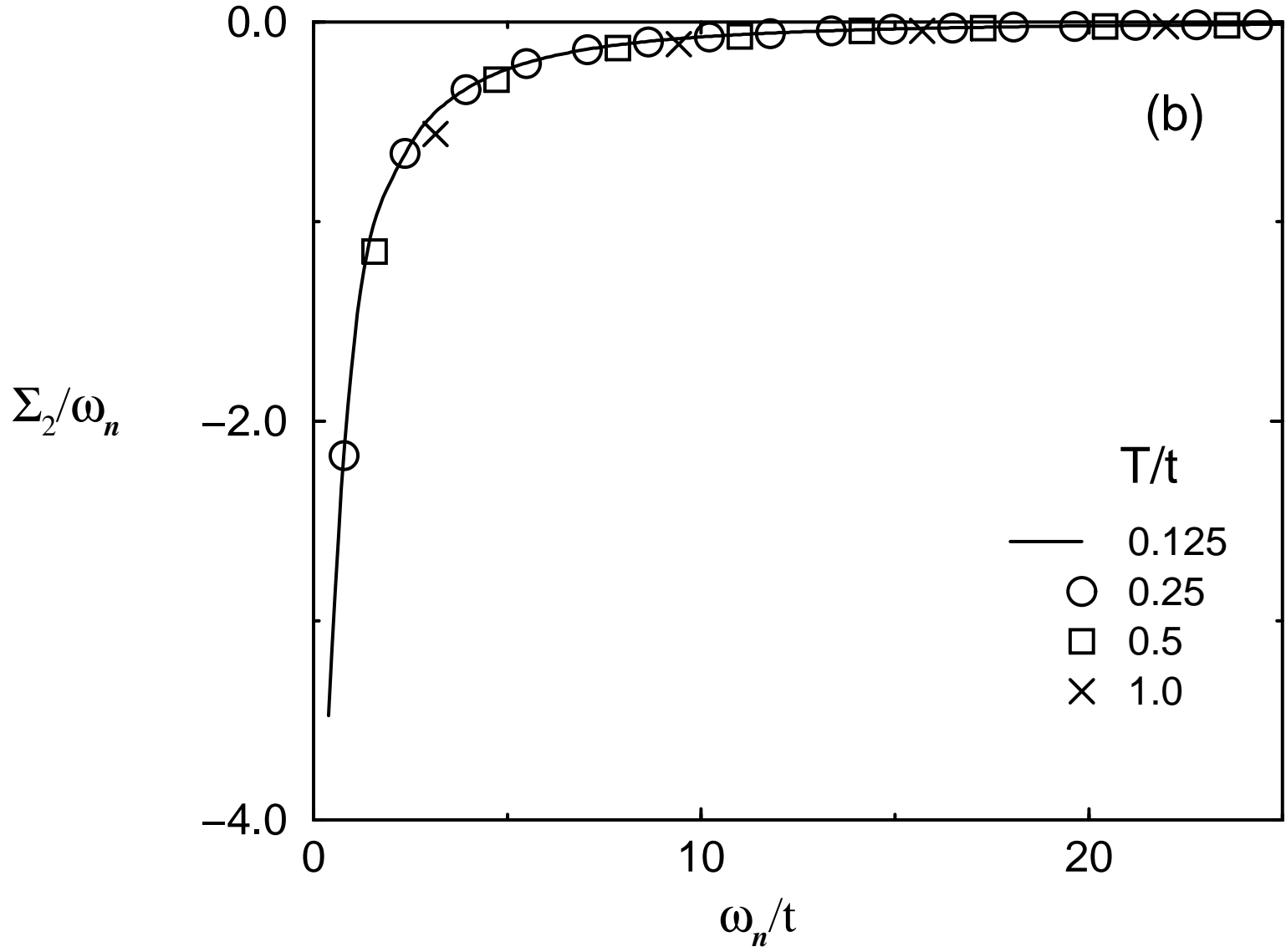
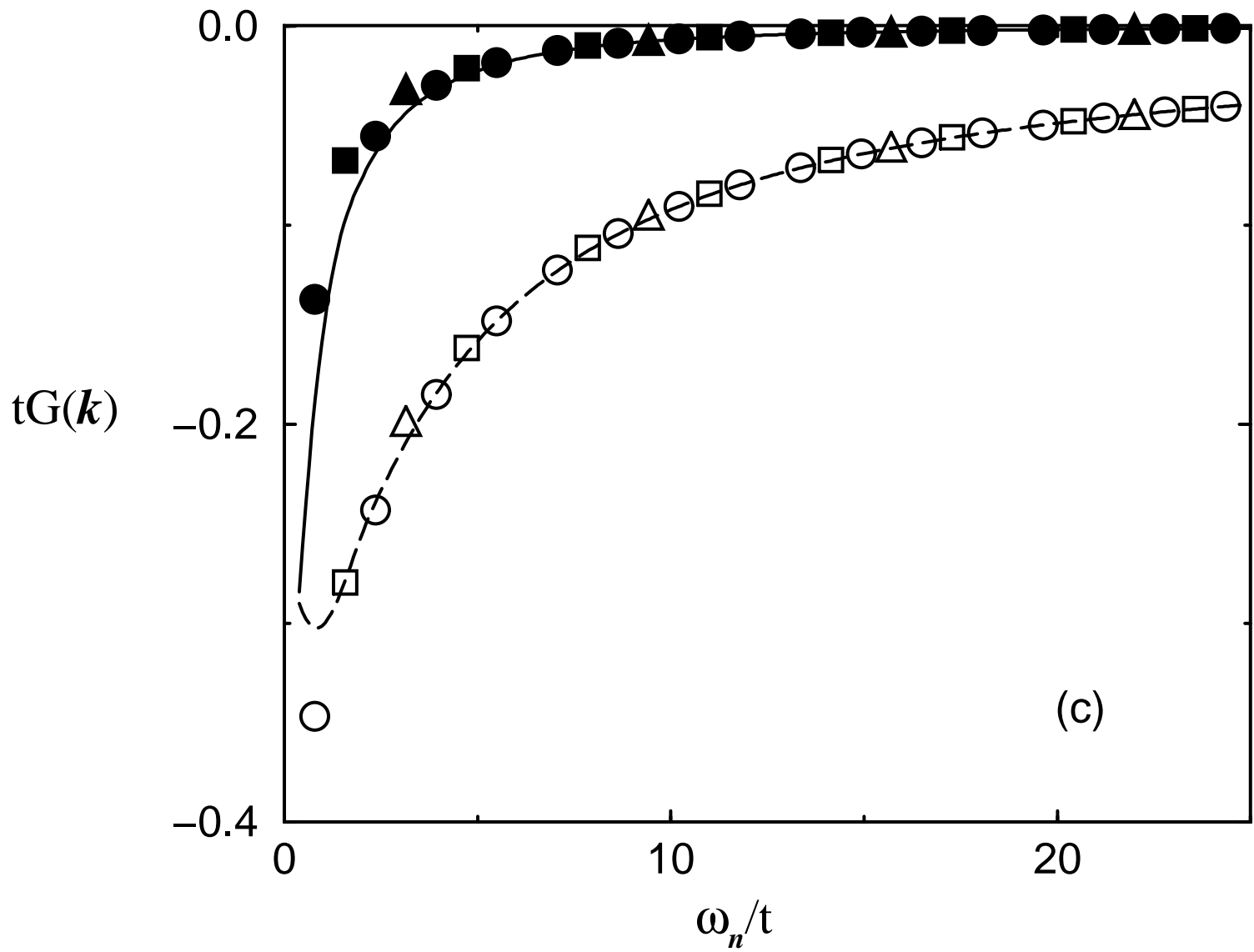
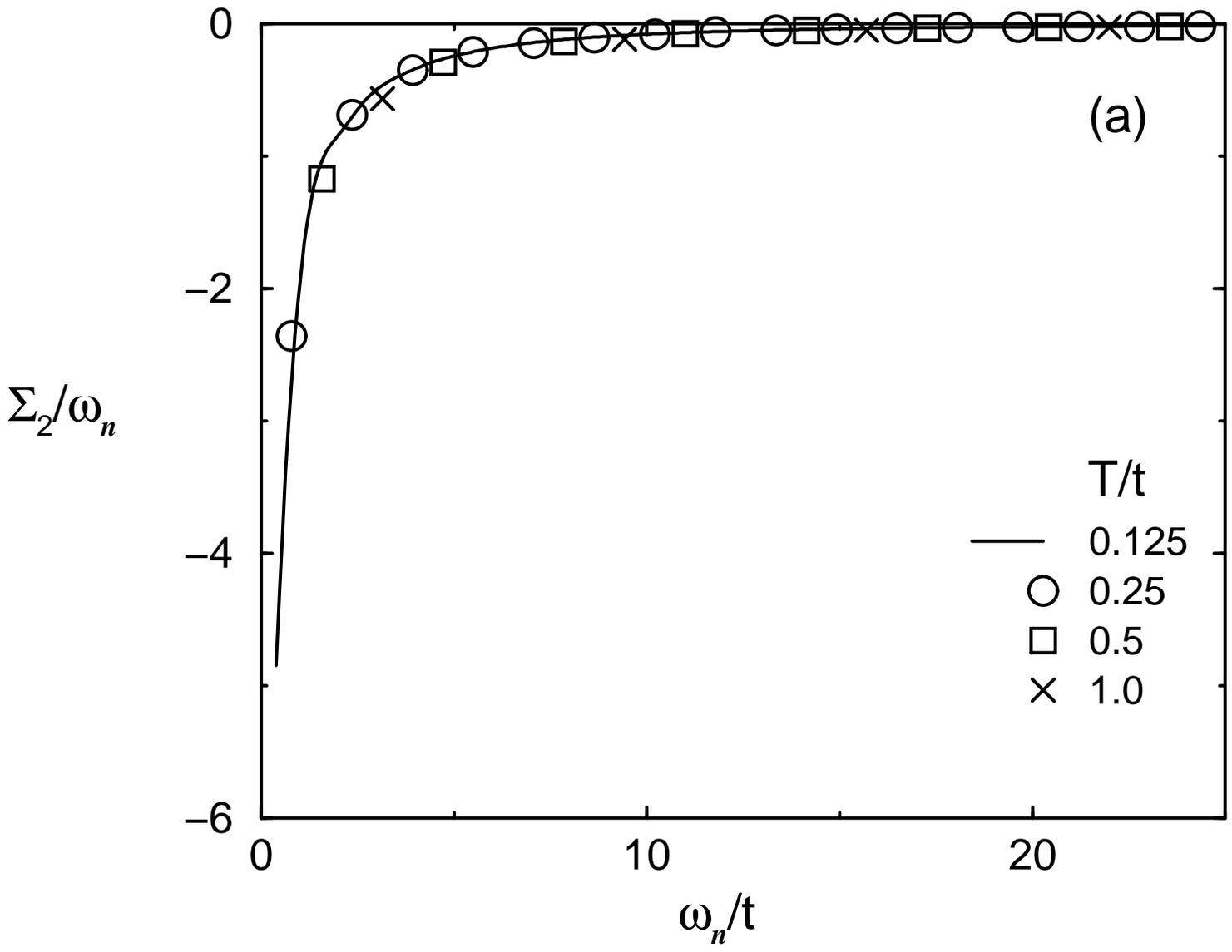
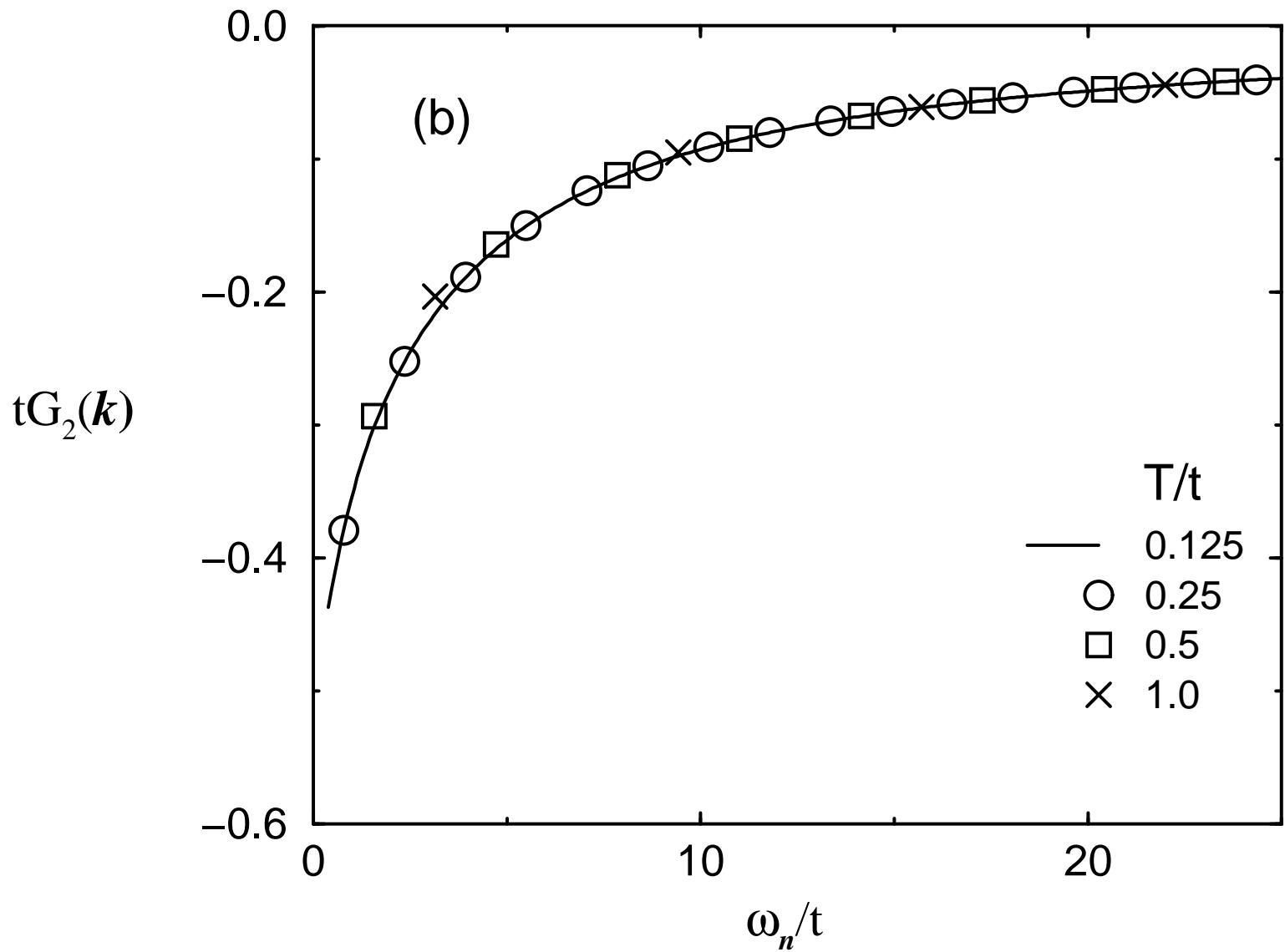
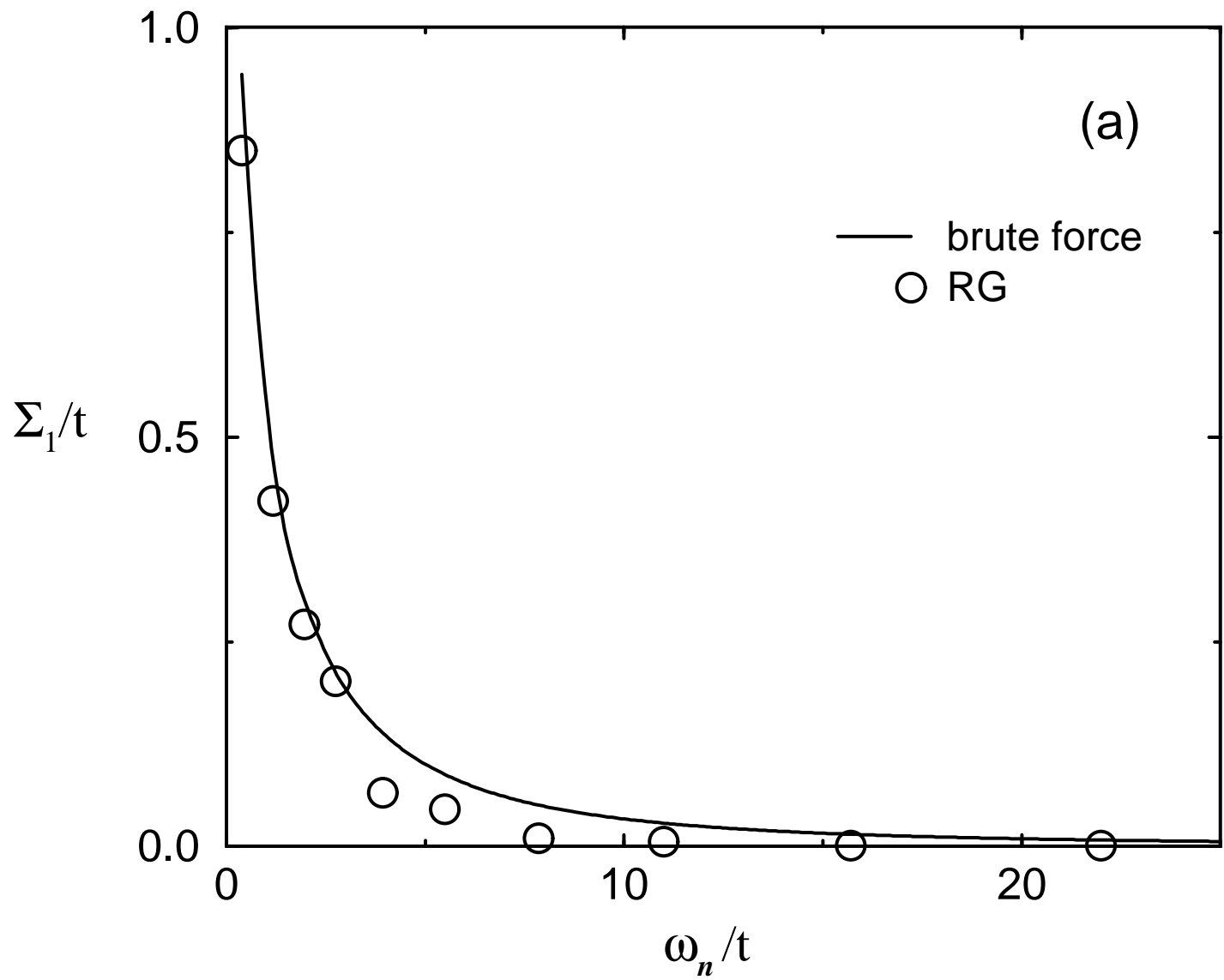
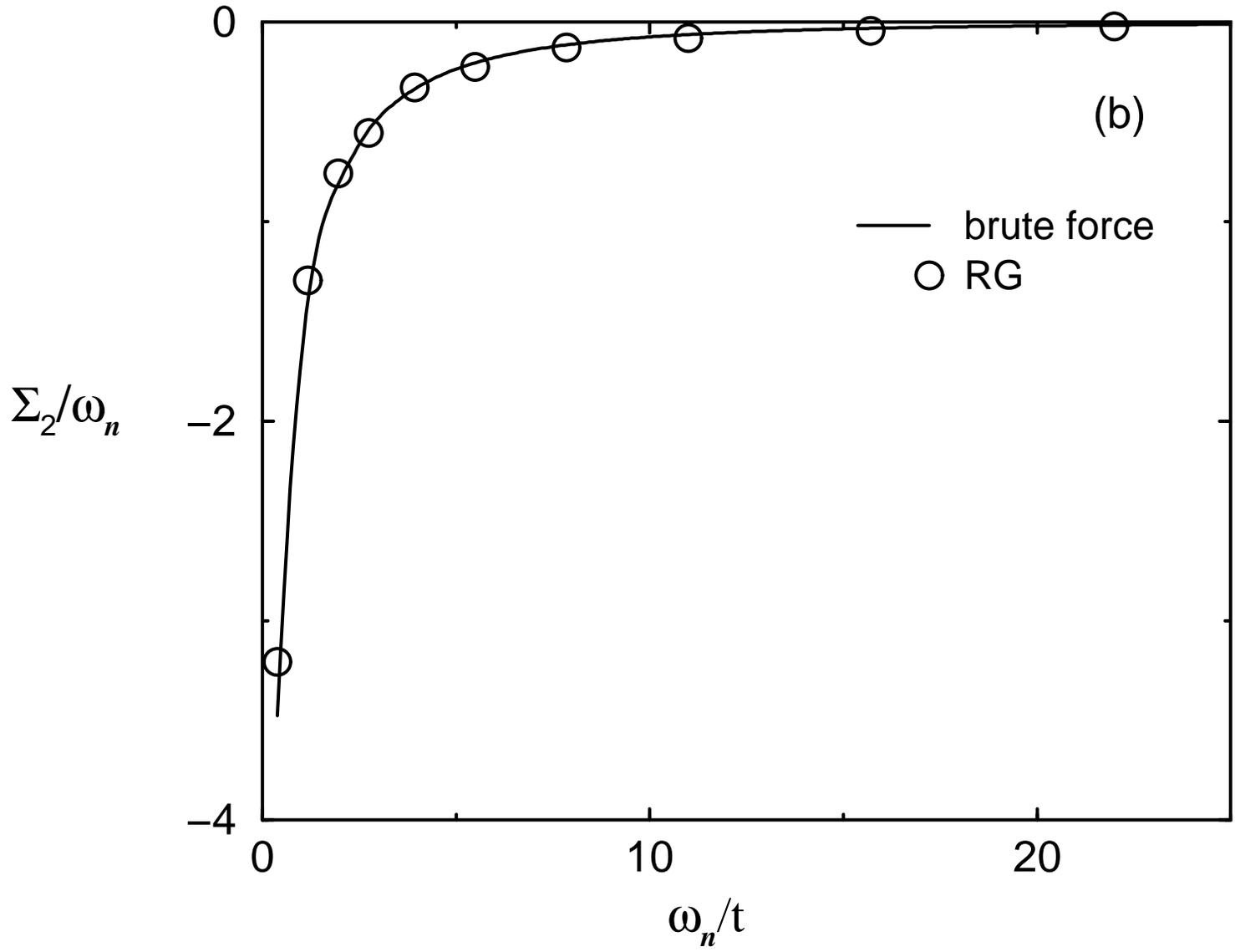
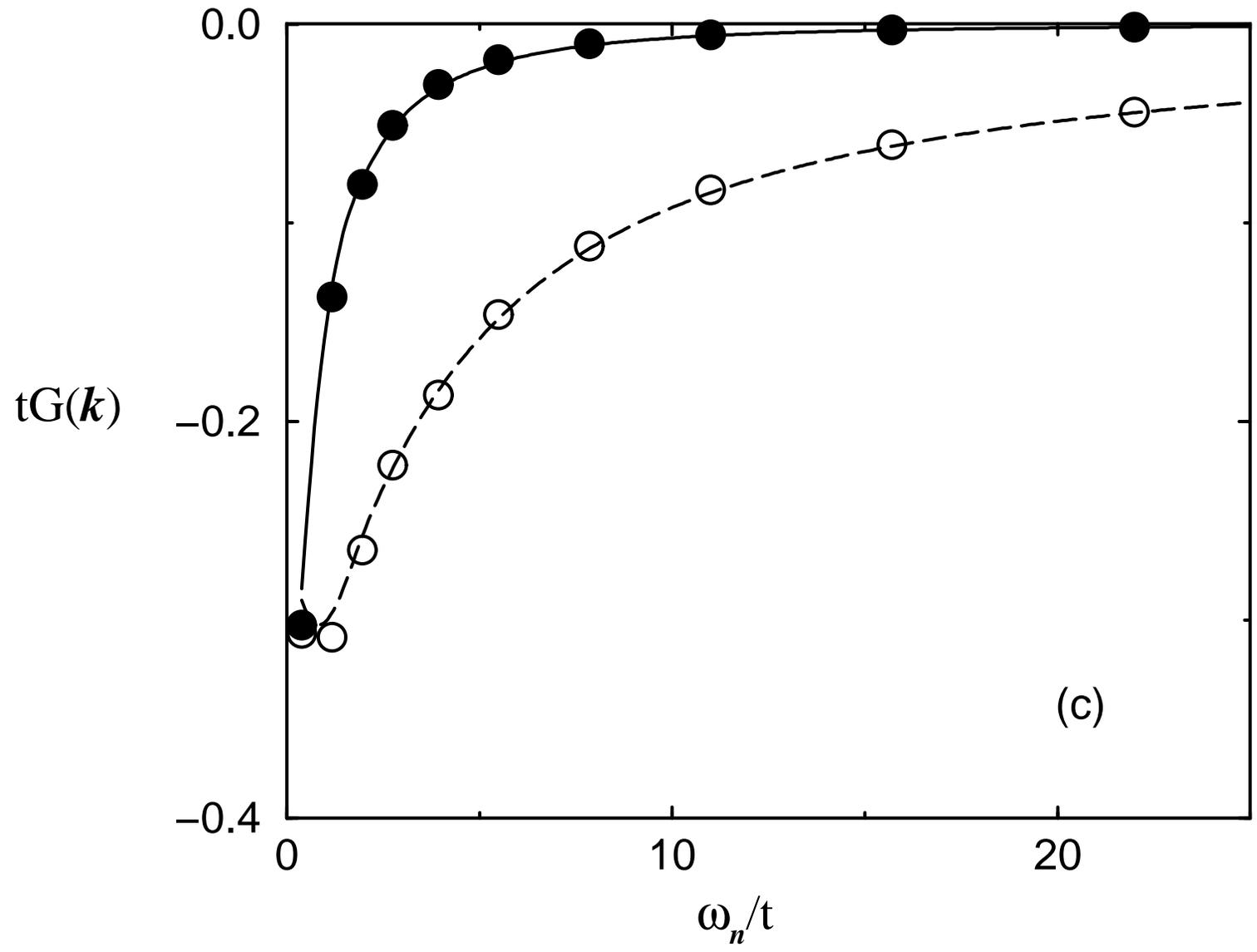
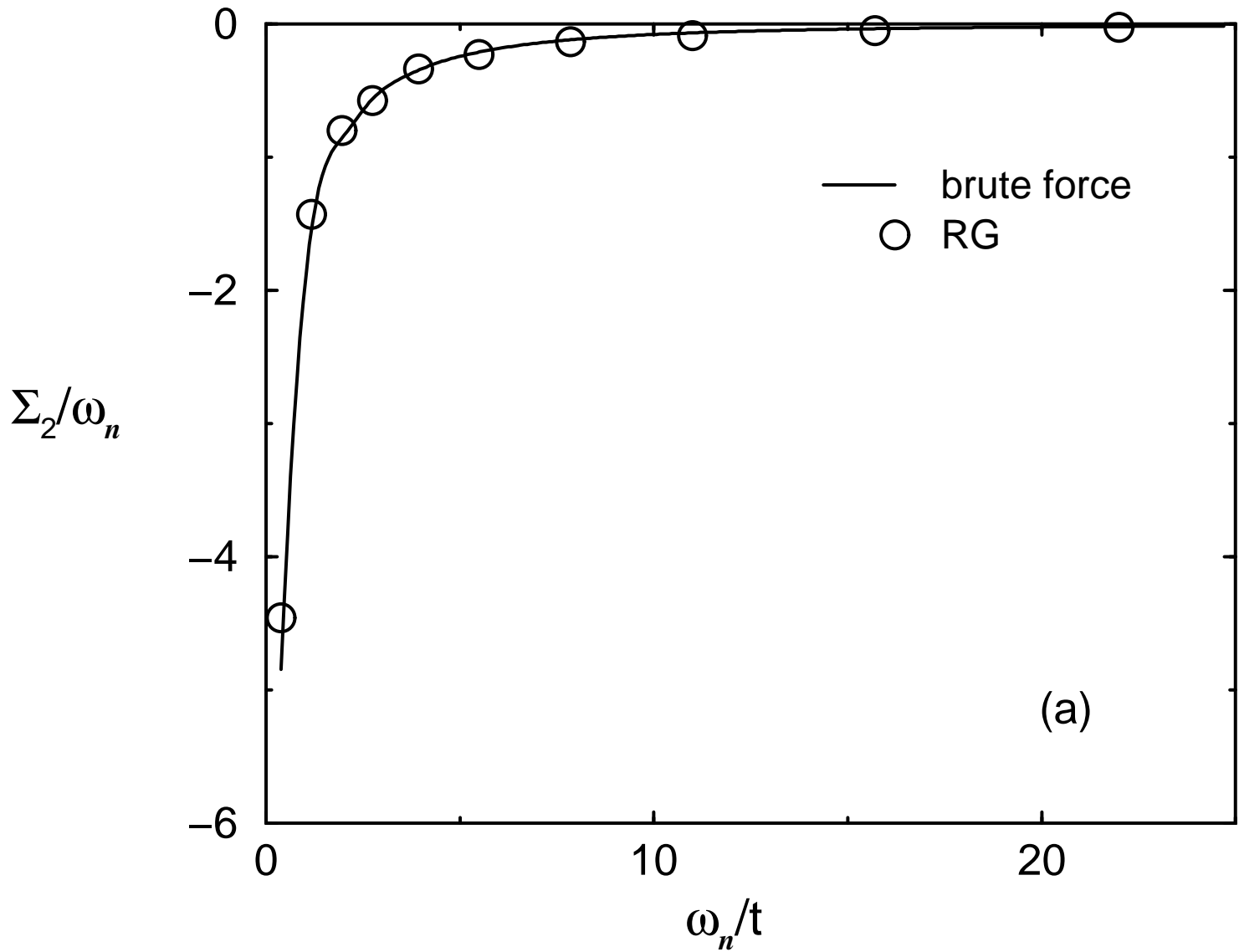
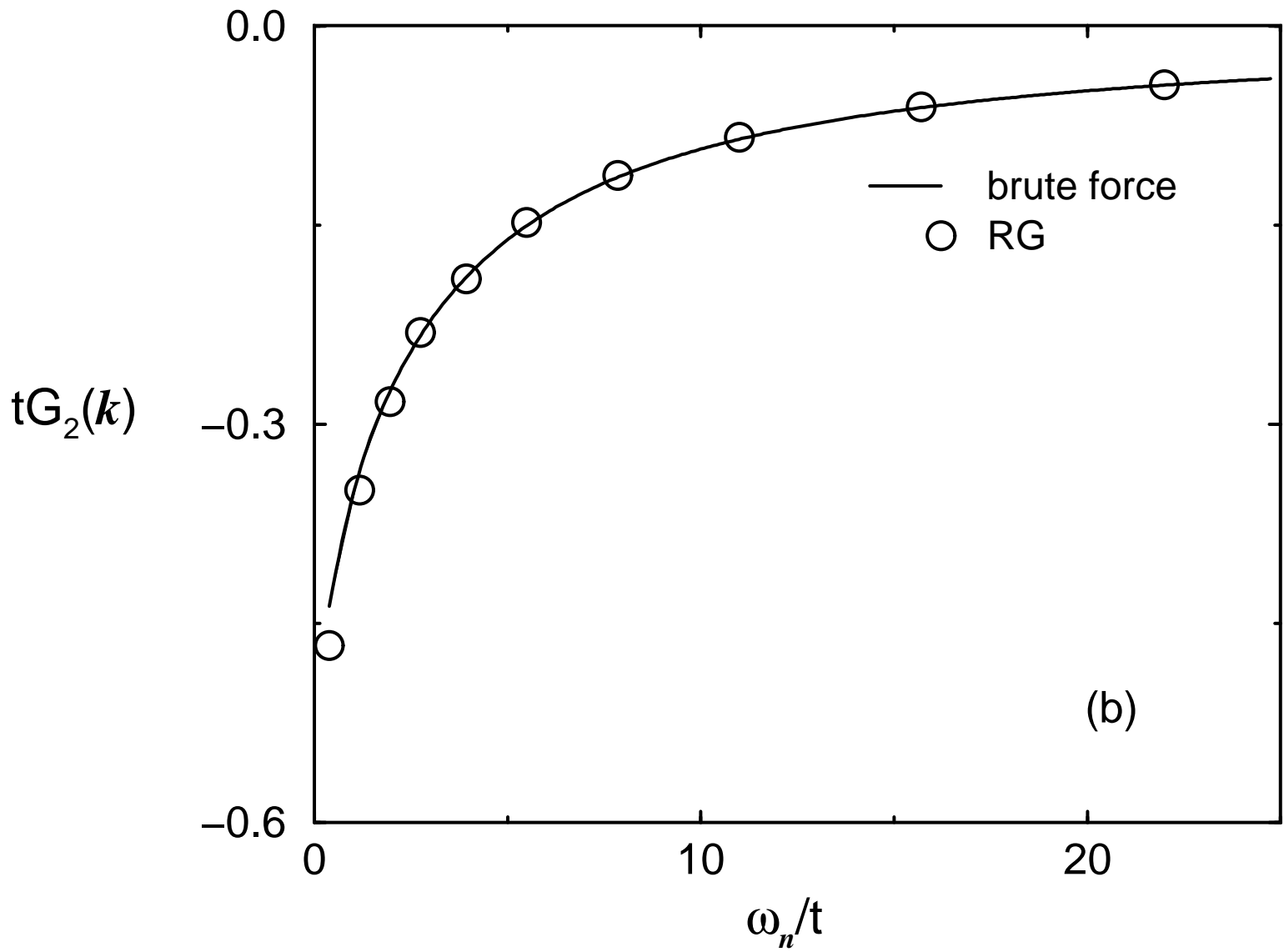
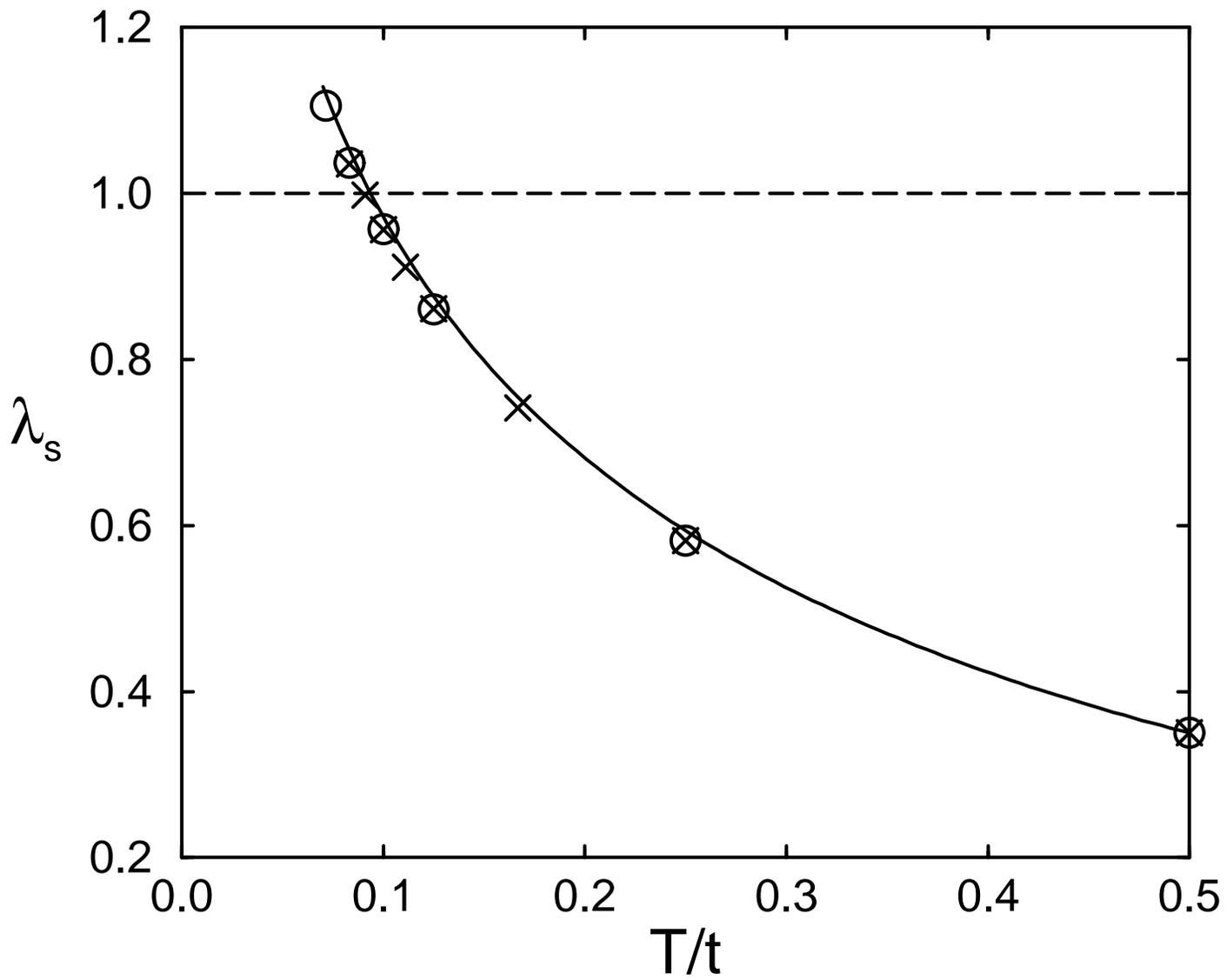
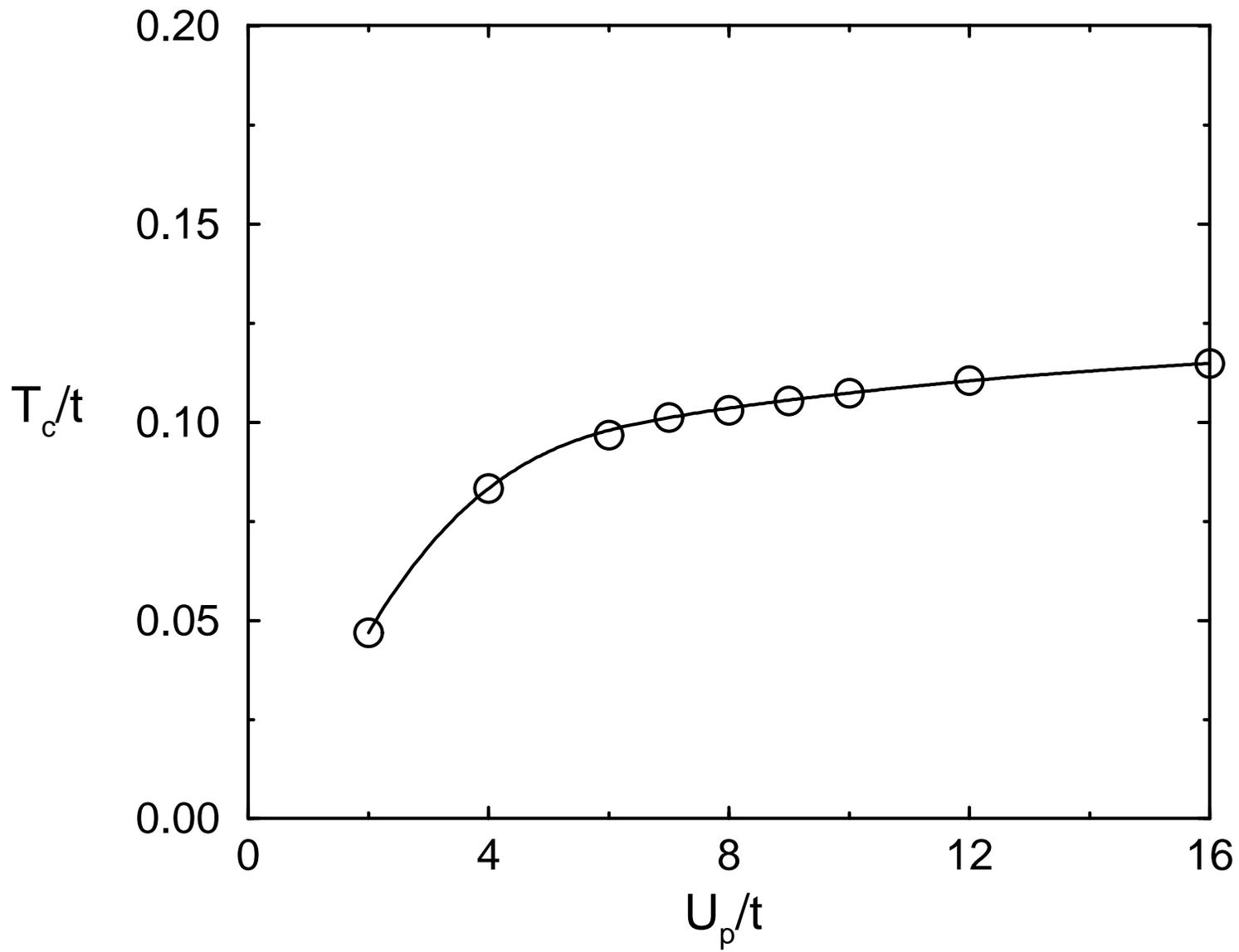
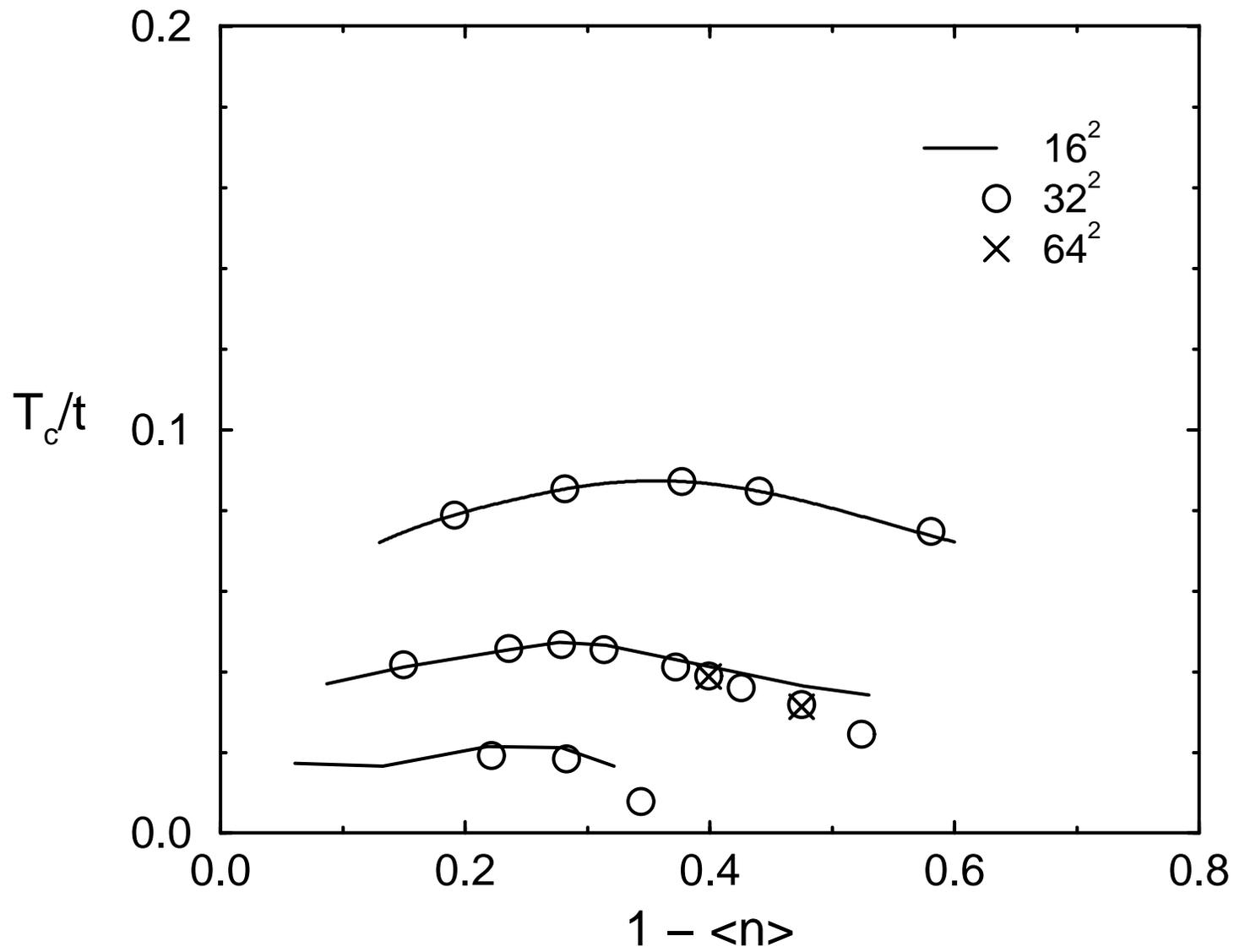
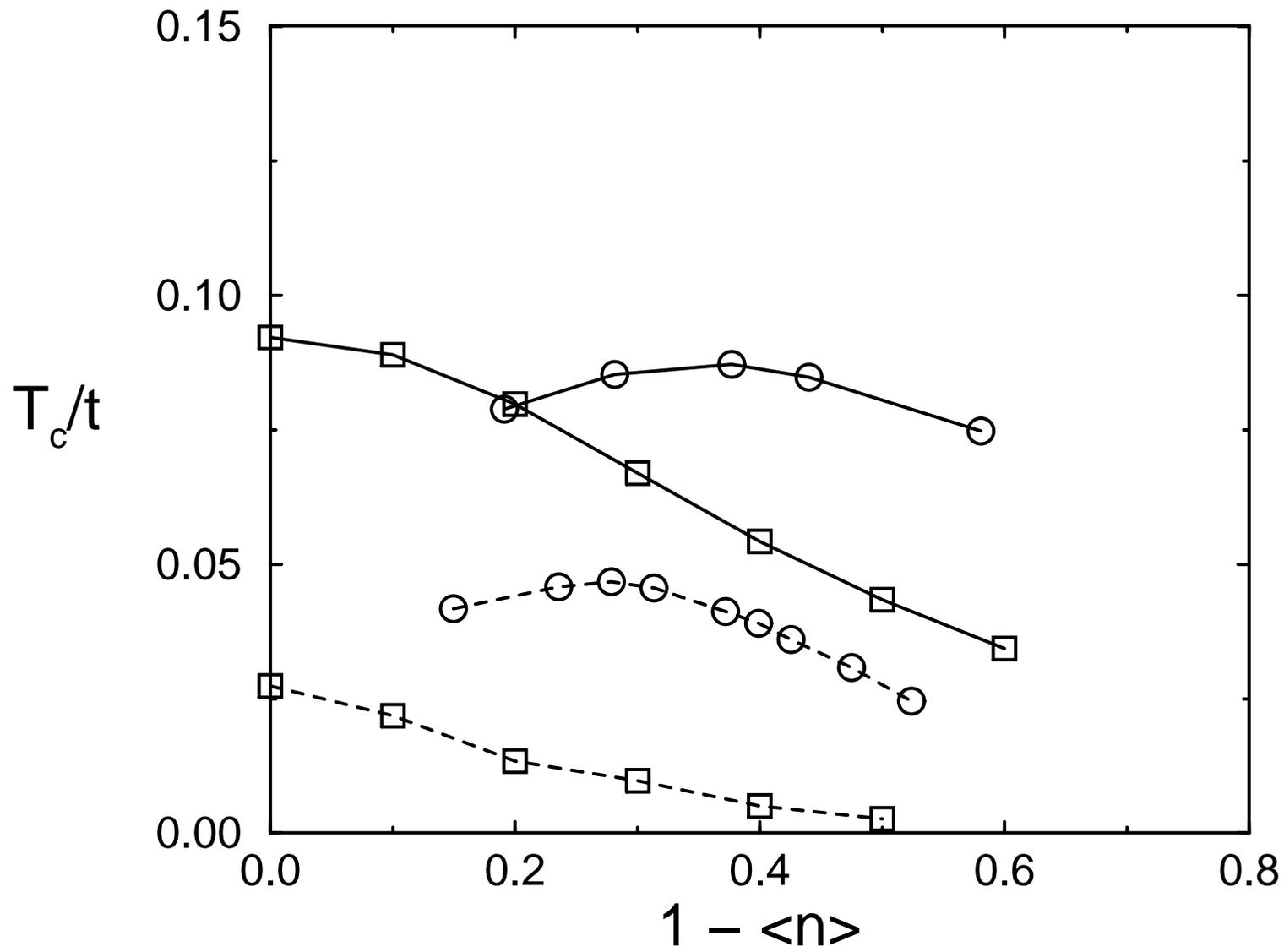
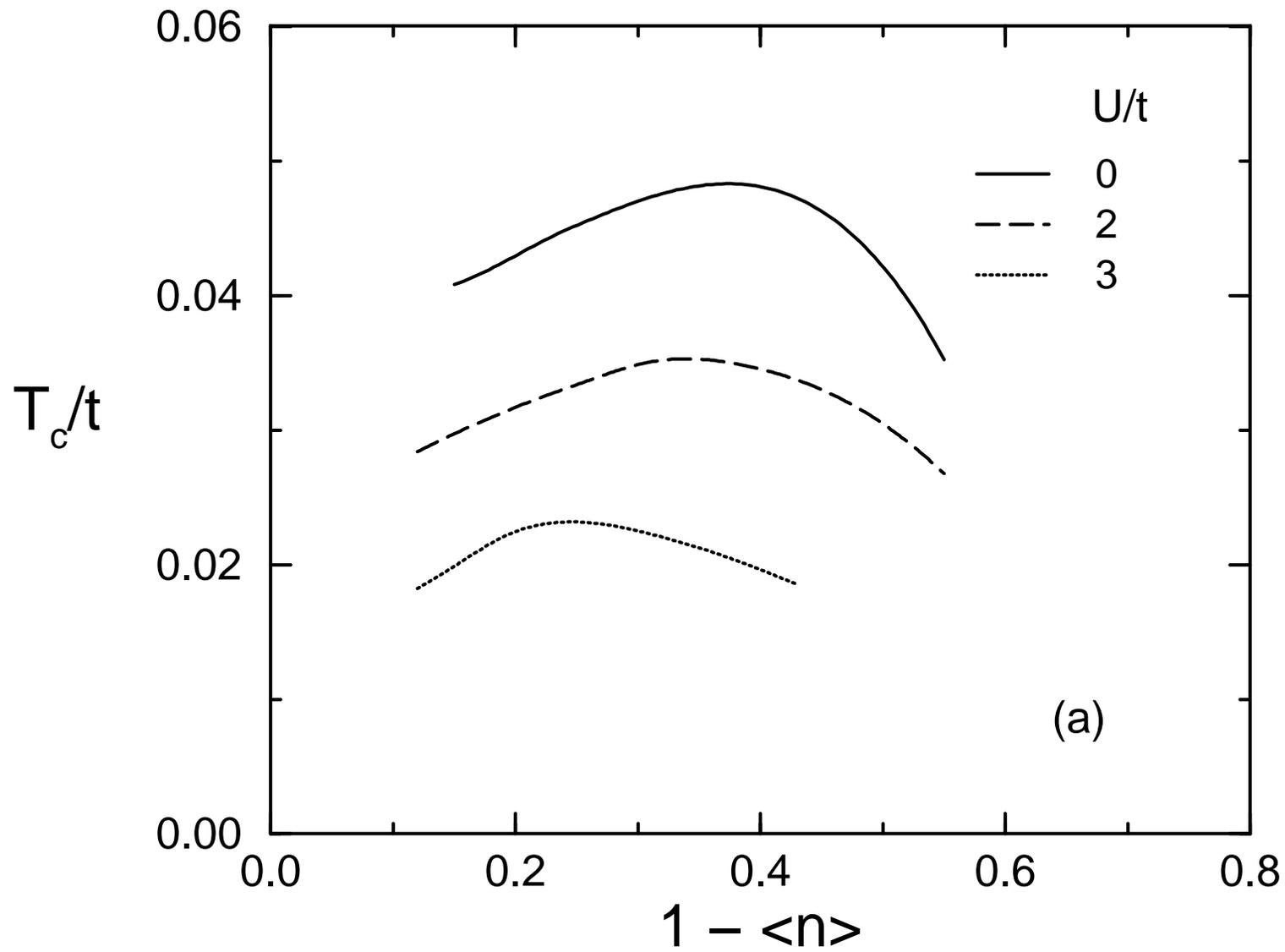
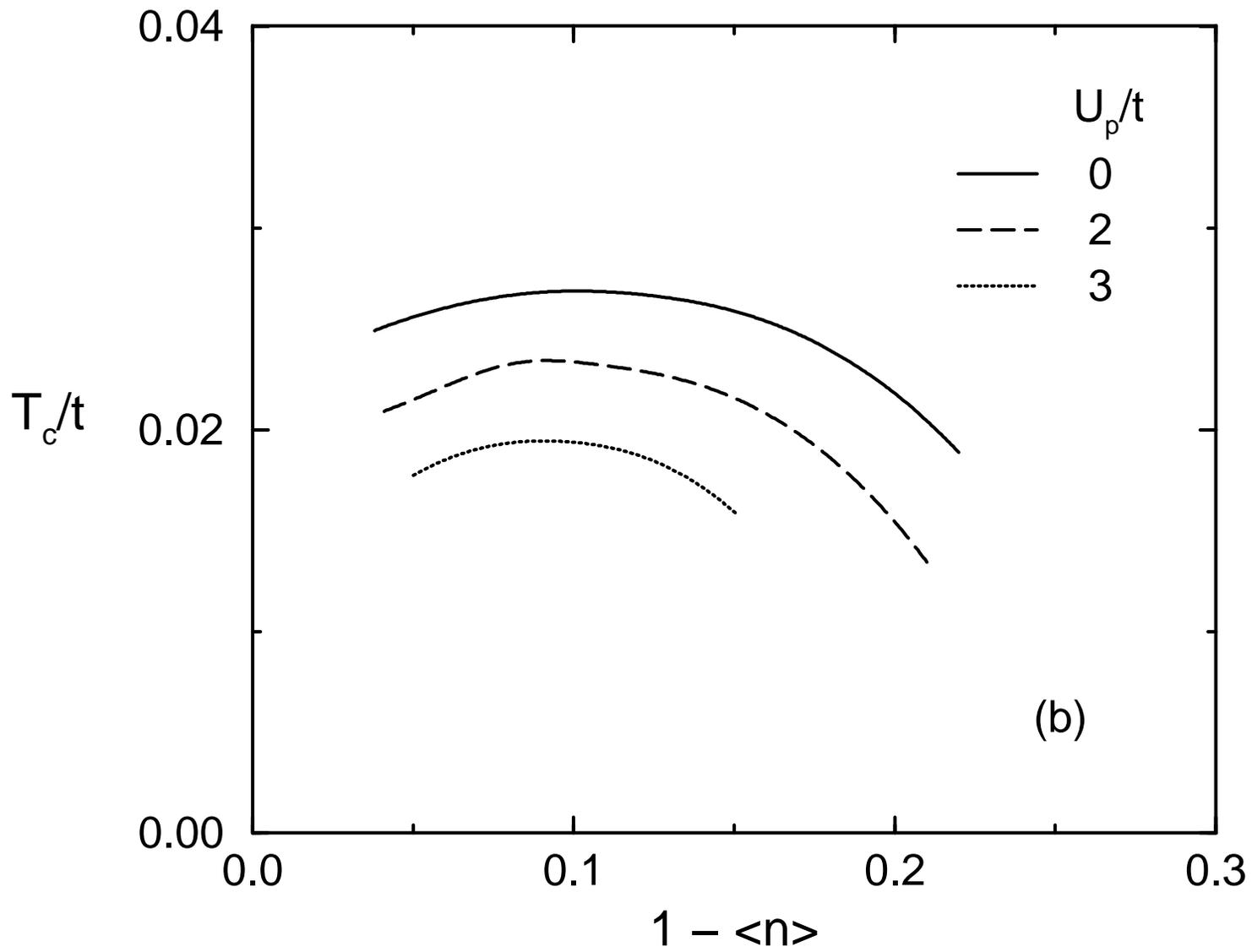
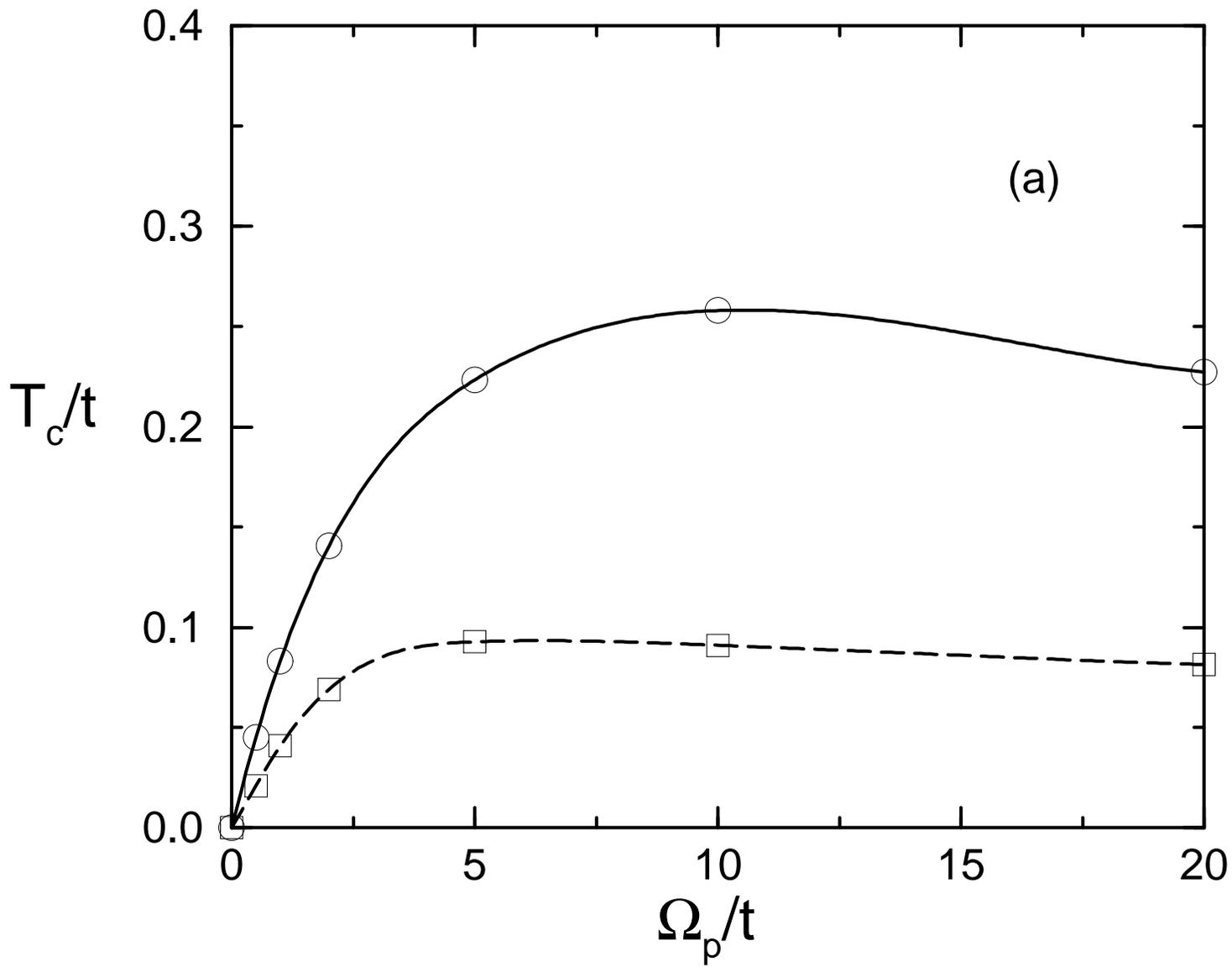
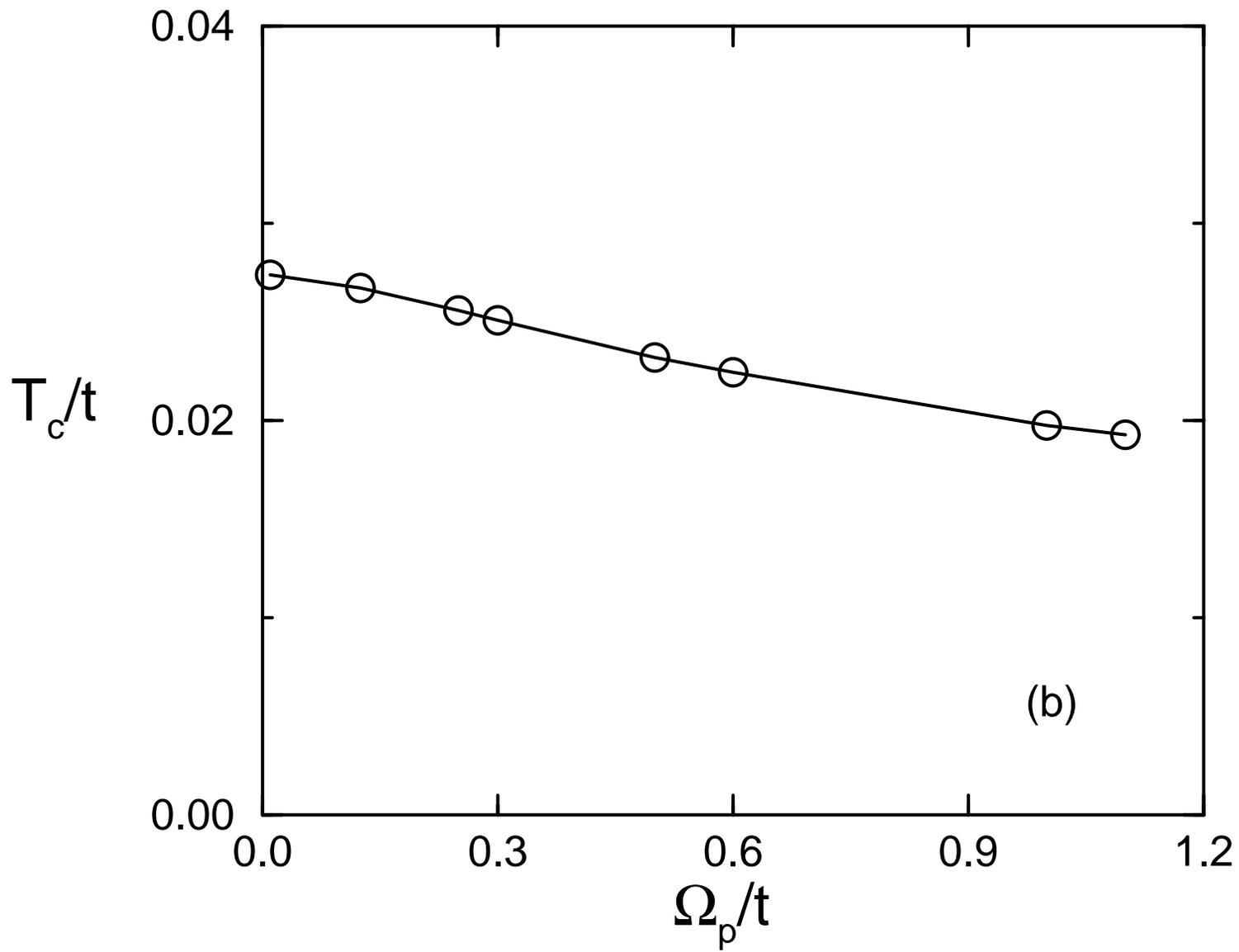
\end{document}